\documentclass{aa}
\usepackage[varg]{txfonts}
\usepackage{natbib}
\bibpunct{(}{)}{;}{a}{}{,}
\usepackage{tabularray}
\usepackage{footnote}
\usepackage{longtable}
\usepackage{lscape}
\usepackage{verbatim}
\usepackage{makecell}
\usepackage{multirow}
\usepackage{booktabs}
\usepackage{soul}
\usepackage{gensymb}
\usepackage{graphicx}
\usepackage{placeins}
\usepackage{ragged2e}
\usepackage{hyperref}

\newcommand\rurl[1]{%
  \href{http://#1}{\nolinkurl{#1}}%
}

\makeatletter
\renewcommand*\aa@pageof{, page \thepage{} of \pageref*{LastPage}}
\makeatother
\begin{document} 

\title{Identifying and characterising the population of hot sub-luminous stars with multi-colour MeerLICHT data}

   \author{P. Ranaivomanana 
          \inst{1,2} \and C. Johnston \inst{1,2} \and P.J. Groot \inst{1,3,4} \and C. Aerts \inst{1,2,5,6} \and R. Lees \inst{4} \and L. IJspeert \inst{2} \and S. Bloemen \inst{1} \and M. Klein-Wolt \inst{1} \and P. Woudt \inst{4} \and E. K\"{o}rding \inst{1} \and R. Le Poole \inst{8} \and D. Pieterse \inst{1}
          } 
   \institute{
   Department of Astrophysics/IMAPP, Radboud University, P.O.Box 9010, 6500 GL Nijmegen, The Netherlands\\
   \email{princy.ranaivomanana@ru.nl}
   \and 
   Instituut voor Sterrenkunde, KU Leuven, Celestijnenlaan 200D, 3001 Leuven, Belgium \and
    South African Astronomical Observatory, PO Box 9, Observatory, 7935, Cape Town, South Africa
    \and Department of Astronomy \& Inter-University Institute for Data
        Intensive Astronomy, University of Cape Town, Private Bag X3, 7701
        Rondebosch, South Africa 
   \and 
    Max Planck Institute for Astronomy, Koenigstuhl 17, D-69117 Heidelberg, Germany
        \and
    Guest Researcher, Center for Computational Astrophysics, Flatiron Institute, 
    162 Fifth Ave, New York, NY 10010, USA
        \and Leiden Observatory, Leiden University, P.O. Box 9513, NL-2300 RA
        Leiden, The Netherlands
    }
    
     \date{Received month day, year; accepted month day, year}
 
  \abstract
   {Colour-magnitude diagrams reveal a population of blue (hot) sub-luminous objects with respect to the main sequence. These hot sub-luminous stars are the result of evolutionary processes that require stars to expel their obscuring, hydrogen-rich envelopes to reveal the hot helium core. As such, these objects offer a direct window into the hearts of stars that are otherwise inaccessible to direct observation.}
    {MeerLICHT is a wide-field optical telescope that collects multi-band photometric data in six band filters (\textit{u, g, r, i, z,} and \textit{q}), whose primary goals are to study transient phenomena, gravitational wave counterparts, and variable stars. We showcase MeerLICHT's capabilities of detecting faint hot subdwarfs and identifying the dominant frequency in the photometric variability of these compact hot stars, in comparison to their \textit{Gaia} DR3 data. We hunt for oscillations, which will be an essential ingredient for accurately probing stellar interiors in future asteroseismology.}
   {Comparative MeerLICHT and \textit{Gaia} colour-magnitude diagrams are presented as a way to select hot subdwarfs from our sample. A dedicated frequency determination technique is developed and applied to the selected candidates to determine their dominant variability using time-series data from MeerLICHT and \textit{Gaia} DR3. We explore the power of both datasets in determining the dominant frequency.}
   {Using the $g-i$ colour, MeerLICHT offers a colour-magnitude diagram that is comparable in quality to that of \textit{Gaia} DR3. The former, however, is more sensitive to fainter objects. The MeerLICHT colour-colour diagrams allow for the study of different stellar populations. The frequency analysis of MeerLICHT and \textit{Gaia} DR3 data demonstrates the superiority of our MeerLICHT multi-colour photometry in estimating the dominant frequency compared to the sparse \textit{Gaia} DR3 data.}
  {MeerLICHT's multi-band photometry leads to the discovery of high-frequency faint subdwarfs. Continued observations tuned to asteroseismology will allow for mode identification using the method of amplitude ratios. Our MeerLICHT results are a proof-of-concept of the capacity of the BlackGEM instrument currently in the commissioning stage at ESO's La Silla Observatory in Chile.}


\titlerunning{Hot sub-luminous stars in the multi-colour MeerLICHT photometry}
\authorrunning{Ranaivomanana et al.}
\maketitle
%
\section{Introduction}
Hot subdwarf O- and B-type stars (sdOs and sdBs, respectively) are a class of sub-luminous, high gravity ($\sim 4.8 <\log g < 5.5$),  post-main-sequence stars that have been relieved of their opaque hydrogen envelopes. As low-mass, core-burning helium stars with a thin hydrogen envelope \citep{Heber2009}, they are located at the extreme horizontal branch (EHB): the bluest end of the horizontal branch in the Hertzsprung-Russell diagram. Decades of research has shown that these objects are ideally suited for providing information for multiple sub-fields of astrophysics. For instance, with their high surface temperatures (20\,000$<T_{\rm eff}<$80\,000 K), sdOB stars are potential sources of ionising ultraviolet radiation \citep[e.g.][]{Dorman1995,Han2007}. Additionally, sdOB stars serve as tracers of the evolutionary pathways of horizontal branch stars in general, and due to their lack of an obscuring hydrogen envelope, they offer a direct probe into the core of core helium-burning objects that followed a standard evolution \citep{Lee1990,Dotter2007,Heber2009,Heber2016}. 

Numerous studies have found that a high fraction of sdOBs exist in close binaries, implying the importance of binary interaction in the creation of these objects \citep[e.g.][]{Maxted2001,Han2002,Han2003,Geier2022}. As such, characterising them and their companions can help constrain the still poorly understood processes of mass transfer and common-envelope evolution \citep[e.g.][]{Toonen2012,Ivanova2013}. Typically, the companions of sdOB stars are white dwarfs, late-type main-sequence stars, or brown dwarfs \citep{Heber2016}. However, studies have also revealed that sdOB stars can be found around early-type Be stars \citep{Wang2021,ElBadry2022,Klement2022,Naze2022}. Furthermore, studies have claimed the detection of planetary \citep{Silvotti2014} and neutron star \citep{Wu2018} companions to sdOB stars, highlighting the importance of sdOB stars to multiple aspects of stellar and planetary evolution. Subdwarf stars are also thought to contribute to the population of close, double white-dwarf binaries capable of generating gravitational waves and type-Ia supernovae \citep{Wang2009,Kupfer2018,Gotberg2020}. In addition to the extrinsic variability caused by their binary companions (e.g. eclipses, reflection effects, and ellipsoidal modulation), sdOB stars have been observed to exhibit both pressure (p) and gravity (g) mode oscillations, with amplitudes ranging from micro- to milli-magnitudes \citep{Kilkenny1997,Charpinet1997,Green2003,Fontaine2003}. The presence of pulsations has enabled asteroseismology to characterise the rotation rates of these stars, determine their internal (chemical) structures, and estimate the mass of the thin hydrogen envelope \citep{Telting2006,Hu2007,Vuckovic2009,Randall2010,VanGrootel2010,Charpinet2011,Pablo2012,Ostensen2014,Zong2016,Ghasemi2017,LynasGray2021}. 

Given the wide diversity of seemingly single and binary subdwarfs, as well as intrinsically variable and apparently non-variable subdwarfs, identifying proper populations has been a challenge over the years. Initial populations were identified and characterised through time-consuming ground-based spectroscopic and photometric campaigns \citep{Green1986,Edelmann2003,Brown2005,Jester2005,Lisker2005,MoralesRueda2006,Green2008,Geier2011}. The advent of space-based missions boosted the discovery of subdwarfs thanks to ultraviolet data, such as those from the Galaxy Evolution Explorer (GALEX) survey \citep{Vennes2011}. This remained the predominant avenue for discovery until the launch of \textit{Gaia} \citep{Gaia2016}, a space mission that has uncovered a substantial number of new hot subdwarfs. For example, \cite{Geier2019} compiled a catalogue of $\sim 40,000$ hot sub-luminous stars from \textit{Gaia} Data Release 2 \citep[DR2; ][]{Gaia2018} and introduced classification schemes for hot sub-luminous star candidates based on colour, absolute magnitude, and reduced proper motion cuts. Their methodology was designed for the identification of new subdwarfs in future surveys. The release of the \textit{Gaia} early Data Release 3 \citep[eDR3; ][]{Gaia2021} increased this catalogue to $ 60,000$ hot sub-luminous stars \citep{Culpan2022}. 

Beyond the initial identification, the most challenging barrier to fully exploiting the binary and asteroseismic potential of sdOB stars remains the time-series characterisation of large samples of sdOB stars. The launch of the {\it Kepler} space mission saw the rapid development of sdOB asteroseismology with the identification of tens of new p- and g-mode pulsating sdB stars as well as a handful of close binaries \citep{Kawaler2010,Ostensen2011,Reed2021}. This success has continued with the launch of the Transiting Exoplanet Survey Satellite \citep[TESS;][]{Ricker2015}, thanks to which dozens more pulsators and close binaries have been observed with high-cadence, high duty-cycle observations \citep{Uzundag2021,Baran2021,Barlow2022}. While the {\it Kepler}, \textit{Kepler-2} (K2), and TESS missions have enabled great advances in the study of subdwarfs, these missions are limited in terms of their position in the sky, time base, or magnitude range. To that end, numerous ground-based photometric surveys, such as the Palomar Transient Factory (PTF), the Zwicky Transient Facility (ZTF), the OmegaWhite Survey, the Asteroid Terrestrial-impact Last Alert System (ATLAS), the All-Sky Automated Survey for Supernovae (ASAS-SN), and the Massive Unseen Companions to Hot Faint Underluminous Stars from SDSS (MUCHFUSS) project, amongst others, have been successful in identifying and characterising (sub-)populations of subdwarfs that show high frequency variability \citep{Ramsay2005,Huber2006,Law2009,Macfarlane2015,Kupfer2017,Heinze2018,Jayasinghe2018,Schaffenroth2018,Coughlin2021,Kupfer2021}. Although such surveys cover a larger area in the sky and have deeper magnitude limits compared to the current generation of space-based photometric missions, they are still limited in terms of colour characterisation and suffer from irregular observing cadences. Posing a further challenge, the characterisation of the sub-population of sdOB stars with wide companions on multiple-year-long orbits requires dedicated radial velocity monitoring  \citep{Vos2017,Vos2020}. 

In order to perform an unbiased and in-depth study of subdwarfs and their sub-populations, it is necessary to use high-precision data from all-sky observations, ideally combining ground- and space-based surveys. Our current work is a step in this direction. We assess the capacity of the MeerLICHT telescope \citep{Bloemen2016,Groot2019a} in the study of subdwarfs. We show that MeerLICHT, and by implication the more powerful BlackGEM instrument currently in its commissioning stage \citep{Bloemen2016,Groot2019b}, will play a large role in discovering and characterising hot subdwarfs in the southern sky. In this paper we present the current MeerLICHT catalogue and explore how it can be used to study the population of sdB stars. Section\, \ref{sec:data_n_obs} describes the MeerLICHT data and their colour properties in comparison with the \textit{Gaia} DR3 data. Our MeerLICHT colour-magnitude diagrams (CMDs) reveal a population of hot, sub-luminous stars with multi-colour observations, whose variability we quantify using a dedicated time-series code (Sect.\, \ref{sec:freqmethod}). The variability results are discussed in Sect.\, \ref{sec:freq_application}, and we conclude the paper in Sect.\, \ref{sec:conclusion}.

\section{Data and observation}\label{sec:data_n_obs}
MeerLICHT is an ongoing southern all-sky survey (with declination $<30$\degree), which started its full-time operations in May 2019 (\citealt{Bloemen2016},\,\citealt{Groot2019a}). The survey aims to study transients, variable stars, and gravitational wave counterparts by co-observing with the Square Kilometre Array (SKA) precursor radio telescope, MeerKAT \citep{Jonas2009}. MeerLICHT is a fully robotic 0.6 metre optical telescope installed at the Sutherland Observatory, South Africa. The telescope offers a field of view of 2.7 square degrees, sampled at 0.56 arcsec/pixel. It provides multi-band photometric data in five filters ($u,g,r,i,z$) similar to those in the Sloan Digital Sky Survey (SDSS, \citealt{Fukugita1996}), with an additional broad filter, referred to as the {$q$ band}. Their passband wavelengths are summarised in Table\,\ref{tab:ml_filters}. 
\begin{table}[]
\centering
\caption{BlackGEM and MeerLICHT passband ranges. \label{tab:ml_filters}}
    \begin{tabular}{cc}
    \hline 
Filter& Wavelength range (nm) \\ \hline
$u$& 350 -- 410 \\
$g$& 410 -- 550 \\
$q$& 440 -- 720 \\
$r$& 563 -- 690 \\
$i$& 690 -- 840 \\
$z$& 840 -- 990 \\ \hline
\end{tabular}   
\end{table}

The data used in this work were retrieved from the so-called MeerLICHT reference images database, part of the forthcoming MeerLICHT Southern All Sky Survey, in all six filter bands. The BlackGEM and MeerLICHT observations were obtained on a regular grid, where 12740 fields tile the available sky. The time coverage across all fields is heterogeneous, largely due to the various surveys that dictate the observing strategy. When MeerLICHT is not paired with MeerKAT, observations are split between trying to build reference images for the MeerLICHT Southern All Sky Survey, follow-up of high interest targets, or a back-up programme. The back-up programme consists of a small subset of 140 fields (with field identification numbers >16000) that are directed towards Local Universe mass over-densities, and often receive higher duty-cycle observations in multiple filters. 

In this work we use a dataset of about 5.5 million objects whose brightness is measured in the $u, g, r, q,$ and $i$ filters. This dataset is extracted from 5137 fields (as of July 2022) that contain at least one observation, with the positions and apparent magnitudes of the object, in each of the five filters. In order to obtain parallaxes for our sources, we cross-match the positions of the sources in our data with that of \textit{Gaia} DR3, considering objects to correspond when they are situated within a radial distance of $\rho < 1$ arcsec. MeerLICHT astrometry is done on the \textit{Gaia} DR2 astrometric frame and shows a systematic rms-error on the frame solution of $\lesssim 50$ milli-arcsec. The positional accuracy on a given source depends on the signal-to-noise ratio of the object but is better than $0.2$ arcsec at the limiting magnitude. Given its larger wavelength coverage, the $q$ band has the most precise photometric measurements for a given integration time. However, the precision does drop off as a function of magnitude, and is further affected by the atmospheric transparency during a given exposure. 

Since the traditional $1/parallax$ method is only valid when the parallax measurements are free from uncertainties \citep{Bailer2015}, we converted parallax into distance (in parsec) using the methodology described in \cite{Bailer2015} and \cite{Astraatmadja2016} and implemented in the {\sc topcat} software \citep{Taylor2005}. In brief, this method estimates the distance based on the mode of a posterior distribution drawn from an exponentially decreasing volume density prior and by assuming the parallax follows a Gaussian distribution \citep{Bailer2015}. As the mode of the posterior depends on a parameter, $L$, known as a scale length, we adopted a value of $L$=1.35 kpc, which provides the best estimate on the distance for true fractional parallaxes $< 2.5$ \citep{Astraatmadja2016}. The mode of the posterior distribution is given by \citep{Bailer2015}\begin{align}
   & & \frac{r^3}{L} - 2r^2 + \frac{\varpi}{\sigma_{\varpi}^2}r - \frac{1}{\sigma_{\varpi}^2} \,=\, 0 \, ,
\end{align}
where $r$ is the estimated distance (pc), L the scale length (pc), and $\varpi$ (arcsec) and $\sigma_{\varpi}$ (arcsec) refer to the measured parallax and its uncertainty, respectively. We only retain sources for which the fractional parallax error is better than 20\% \citep[i.e. $\sigma_{\varpi}/ \varpi <0.2$;][]{Bailer2015}. Additionally, to make a more robust selection, \textit{Gaia} quality flags are applied to the data, including renormalised unit weight error (RUWE) $<$ 1.4 \citep{Lindegren2021} and fidelity\_v2 $>$ 0.5 (e.g. \citealt{Rybizki2022}). We use these quality cuts to ensure that our sample consists of stars with reliable astrometric solutions. After applying these cuts, we are left with $\sim$4.6 million cross-matched objects, which are used to construct our CMDs and colour-colour diagrams discussed in Sect. \ref{sec:cmd}.

\subsection{MeerLICHT photometry}
In this section we briefly discuss the extraction of MeerLICHT photometry, compare it to \textit{Gaia} photometry, and use it to construct diagnostic diagrams. As previously mentioned, the astrometric solution for a given observed MeerLICHT image is calculated using the \textit{Gaia} DR2 astrometric frame \citep{Lindegren2018}. All sources in a given image are detected using SourceExtractor \citep{Bertin1996}. Subsequently, the flux of each source is measured using optimal photometry based on position-dependent point spread function shape (\citealt{Naylor1998}; \citealt{Horne1986}). All of the flux measurements are then associated with existing sources using a 1 arcsec association radius. If  there is no existing source within 1 arcsec, then a new entry will be created in the MeerLICHT database. At set dates, a reference image for each filter has been created for each field in the database by performing a median combination of all available images in that filter. In addition to improving the overall signal-to-noise ratio in the reference image, most slow moving transients such as asteroids, comets, and other faint Solar System objects are removed via this process. 

\begin{figure*}
  \centering
 \begin{tabular}{c}
  \includegraphics[trim={0cm 1.75cm 0cm 0cm }, width=0.95\linewidth]{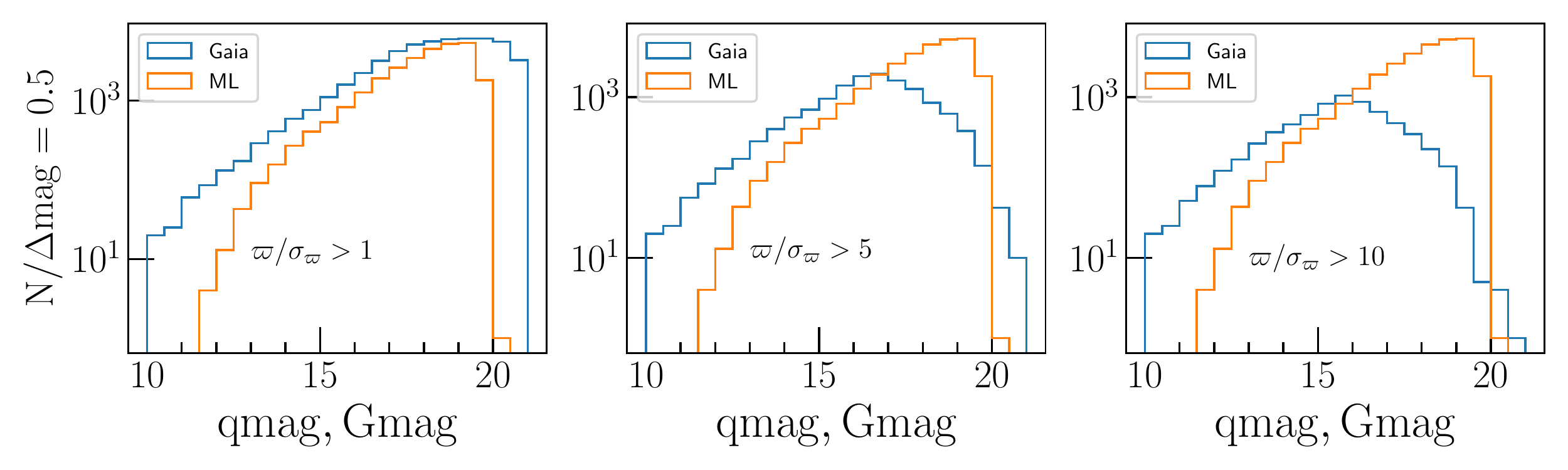}\\
    \includegraphics[trim={0cm 0cm 0cm 0.75cm }, width=0.95\linewidth]{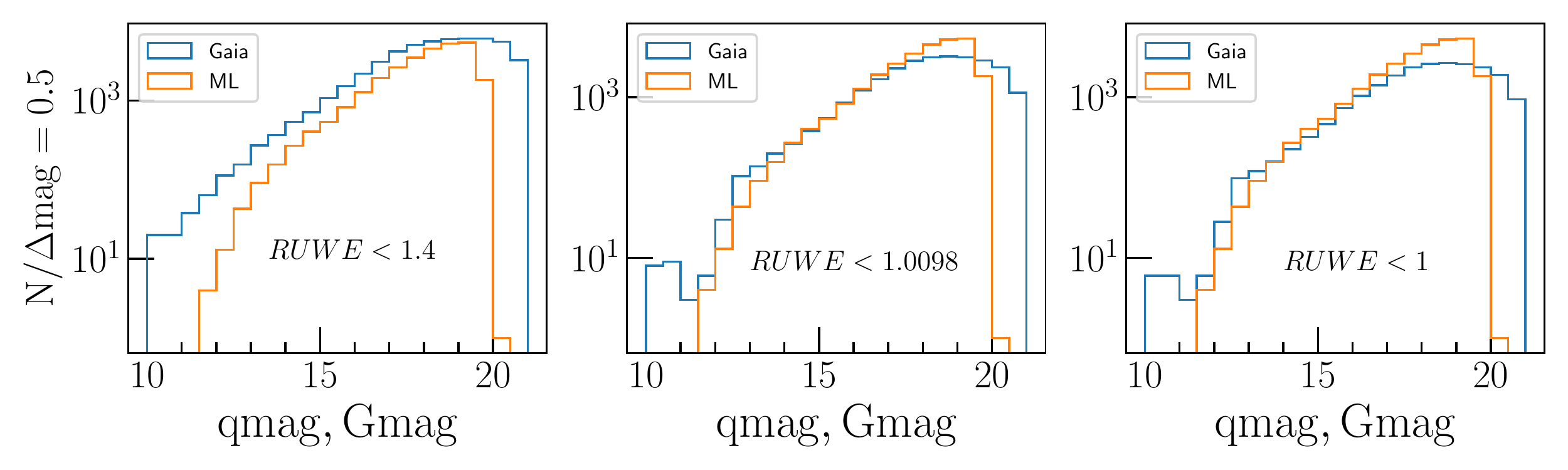}
\end{tabular}
\caption{
Magnitude distribution of the MeerLICHT (ML) $q$ band and the \textit{Gaia} G band in a field with a galactic latitude above 20 $\degree$ ($b = 21.08 \degree$), using a magnitude bin width of $\Delta mag = 0.5$ mag. The y-axis represents the number of data points per magnitude bin, and the x-axis corresponds to MeerLICHT's $q$-band magnitude (qmag) and \textit{Gaia} G-band magnitude (Gmag). The parallax over parallax error ($\varpi/\sigma_{\varpi}$) and the RUWE cuts are applied independently to the \textit{Gaia} data. The value in the middle bottom panel corresponds to the median of the RUWE.
}\label{fig:ml_vs_gaia_field_b_gt20}
\end{figure*}

\begin{figure*}
  \centering
 \begin{tabular}{c}
  \includegraphics[trim={0cm 1.75cm 0cm 0cm },width=0.95\linewidth]{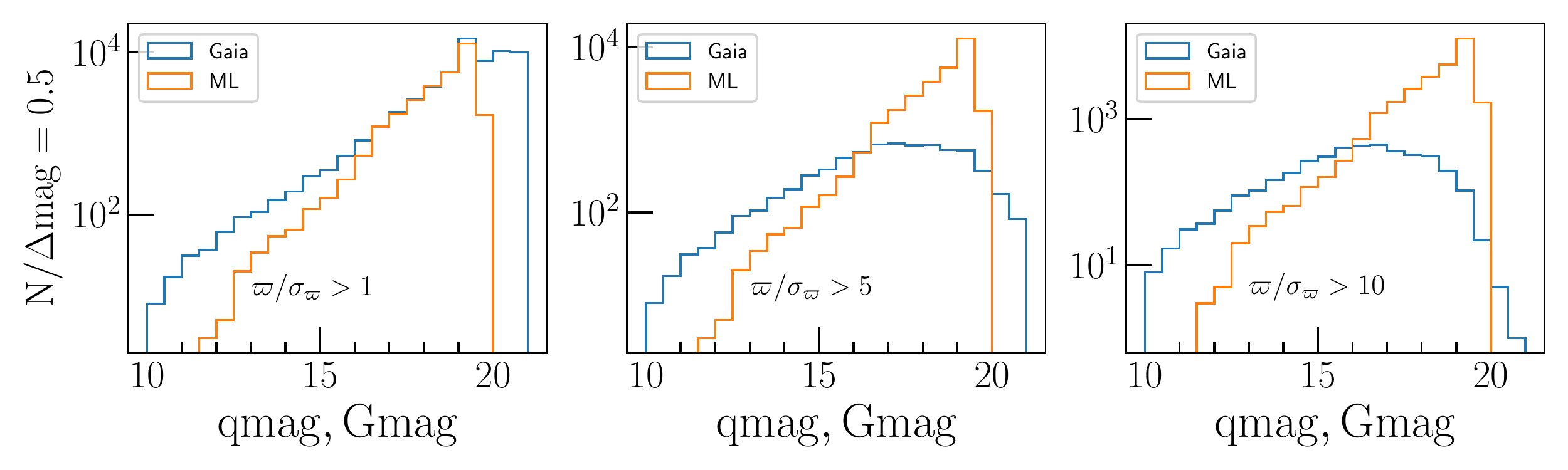}\\
    \includegraphics[trim={0cm 0cm 0cm 0.75cm }, width=0.95\linewidth]{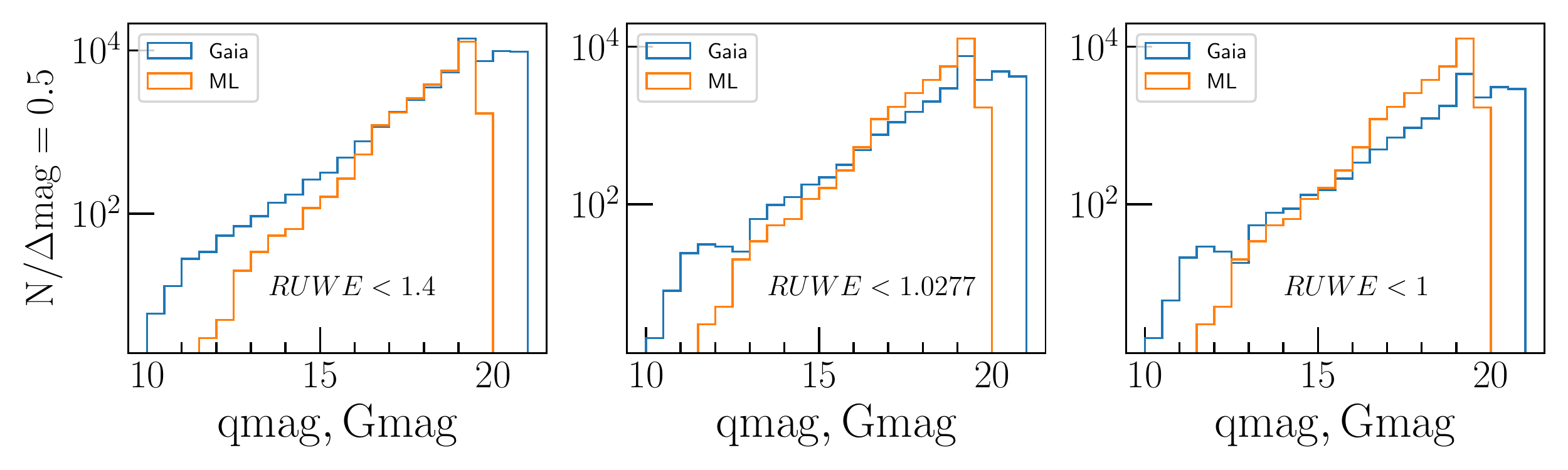}
\end{tabular}
\caption{Same as Fig.  \ref{fig:ml_vs_gaia_field_b_gt20}, but for a field centred at a galactic latitude below $20 \degree$ ($b=-45.06 \degree$). 
}\label{fig:ml_vs_gaia_field_b_lt20}
\end{figure*}

Extracting only stellar sources in the reference images with the $q$ band gives us about 150 million stellar sources (as of July 2022), across $\sim 11000$ fields. It should be noted that these are high fidelity sources, meaning they have  a star class probability above 0.8 (Source Extractor classification flag; star =1; extended source=0). To compare the distribution of observed MeerLICHT sources with that of \textit{Gaia}, we randomly selected two MeerLICHT fields centred at a Galactic latitude $|b| > 20 \degree$. We made a box query of the \textit{Gaia} DR3 archive\footnote[1]{\url{https://gea.esac.esa.int/archive/}} using the corners of each MeerLICHT field. Figures~\ref{fig:ml_vs_gaia_field_b_gt20} and \ref{fig:ml_vs_gaia_field_b_lt20} illustrate the distributions of both datasets with $b=21.08 \degree$ and $b=-45.06 \degree$, respectively, using a magnitude bin width of 0.5 mag. We applied several cuts to the \textit{Gaia} data using the parallax over parallax error ($\varpi/\sigma_{\varpi}$) and the RUWE parameters to quantify their effects on the number of detected sources. We note that the RUWE parameter is expected to be around 1 for single sources \citep{Lindegren2018ruwe}, which drives us to use its median value in the cuts (middle bottom panels of Figs. \ref{fig:ml_vs_gaia_field_b_gt20} and \ref{fig:ml_vs_gaia_field_b_lt20}). Both figures show that the parallax error cuts are more sensitive to fainter objects, while lower values of RUWE most likely remove bright, non-stellar objects. It is worth noting that we applied these two cuts independently, such that when RUWE cuts are applied, there is no restriction on the parallax error, and vice versa. 

We find general agreement between the MeerLICHT and \textit{Gaia} sources within magnitudes between 14 and 19, with ${\rm RUWE}$ values $\lesssim 1.0$. For higher values of parallax error and RUWE cuts, we lose a significant number of \textit{Gaia} data points, especially due to the parallax error cuts.  Furthermore, both figures show that for fainter MeerLICHT objects with $q$-band magnitude $\gtrsim 17$ mag, we are likely to find fewer \textit{Gaia} object matches with good astrometric solutions, $\varpi/\sigma_{\varpi} > 5$. We note that the significant drop in the number of data points in the brighter and fainter ends of the distributions represents MeerLICHT's limiting magnitude ($\sim 20$ mag) and saturation limit ($\sim 12$ mag) in the $q$ band, respectively.

\subsection{Colour-magnitude diagrams}\label{sec:cmd}
\begin{figure*}
  \centering
 \begin{tabular}{c}
\includegraphics[width=1\linewidth]{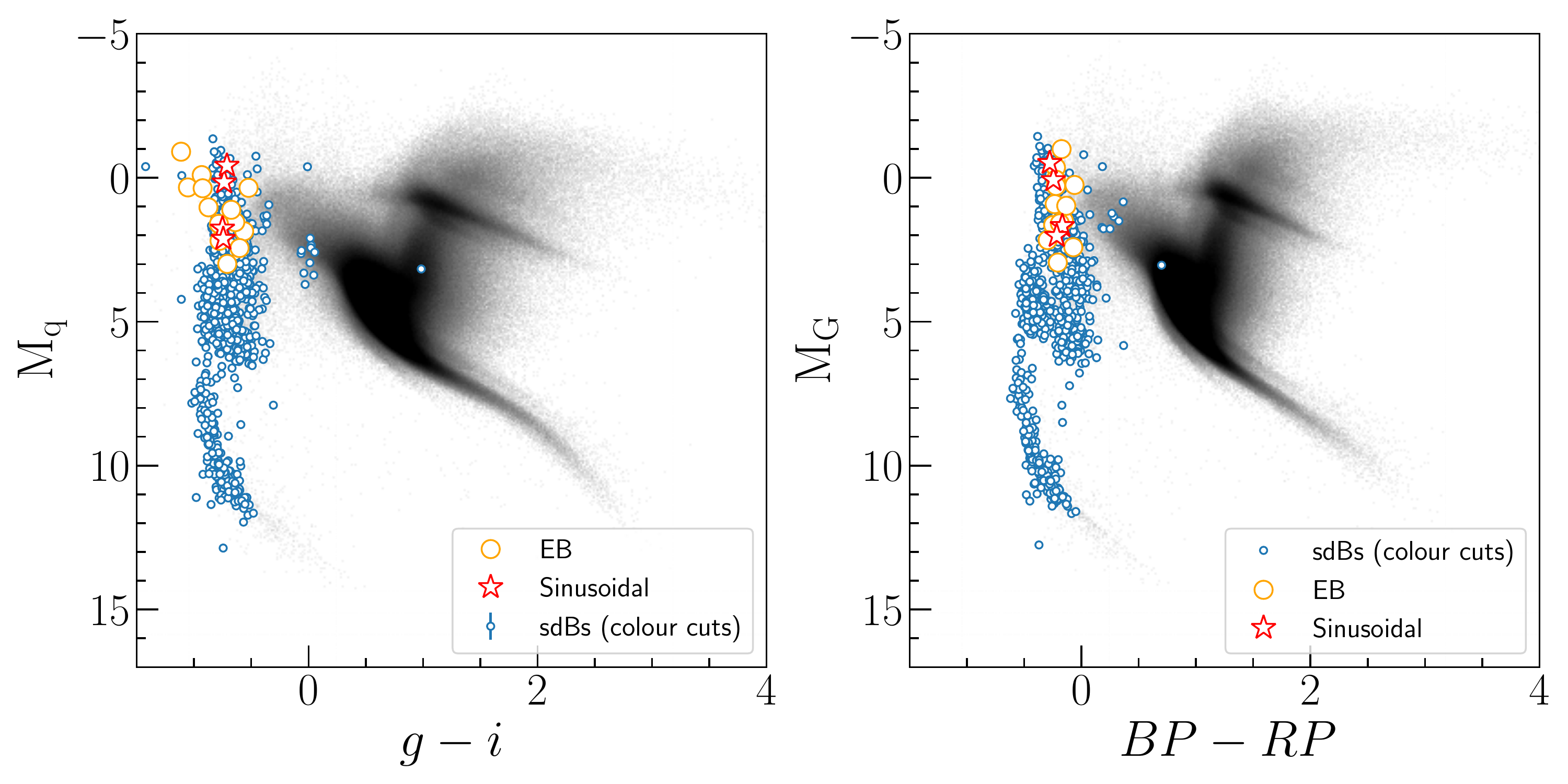}    
\end{tabular}
\caption{CMDs of both MeerLICHT (left) and \textit{Gaia} DR3 (right) data. The blue data points represent the selected sdB candidates from the colour classification schemes. The open orange circles represent EBs identified from the sdB candidates, while the red stars correspond to candidates with sinusoidal variations. Even though white dwarfs are also selected, we do not exclude them from our analysis as they could be relevant to the study of sdB--white dwarf binary systems.  
}\label{fig:cmd_ml_gaia}
\end{figure*}

We constructed our CMDs for the cross-matched and filtered list of $\sim 4.6$ million sources. The resulting CMDs for MeerLICHT and \textit{Gaia} are shown in Fig.~\ref{fig:cmd_ml_gaia}. The left panel depicts the MeerLICHT $(g-i)$ colour versus the $q$-band absolute magnitude, while the right panel represents the \textit{Gaia} $(BP-RP)$ versus the G-band absolute magnitude.
The MeerLICHT and \textit{Gaia} CMDs show a high degree of similarity, which illustrates the power of the MeerLICHT multi-colour photometry to define and extract stellar populations over a wide magnitude range. The CMDs of both datasets clearly reveal the longest stellar evolution phases, such as the main sequence, the giant branch, the EHB, and the white dwarf cooling track. 

\subsection{sdB candidate selection}\label{sec:sdb_selection}
\begin{figure*}
  \centering
  \includegraphics[width=1\linewidth]{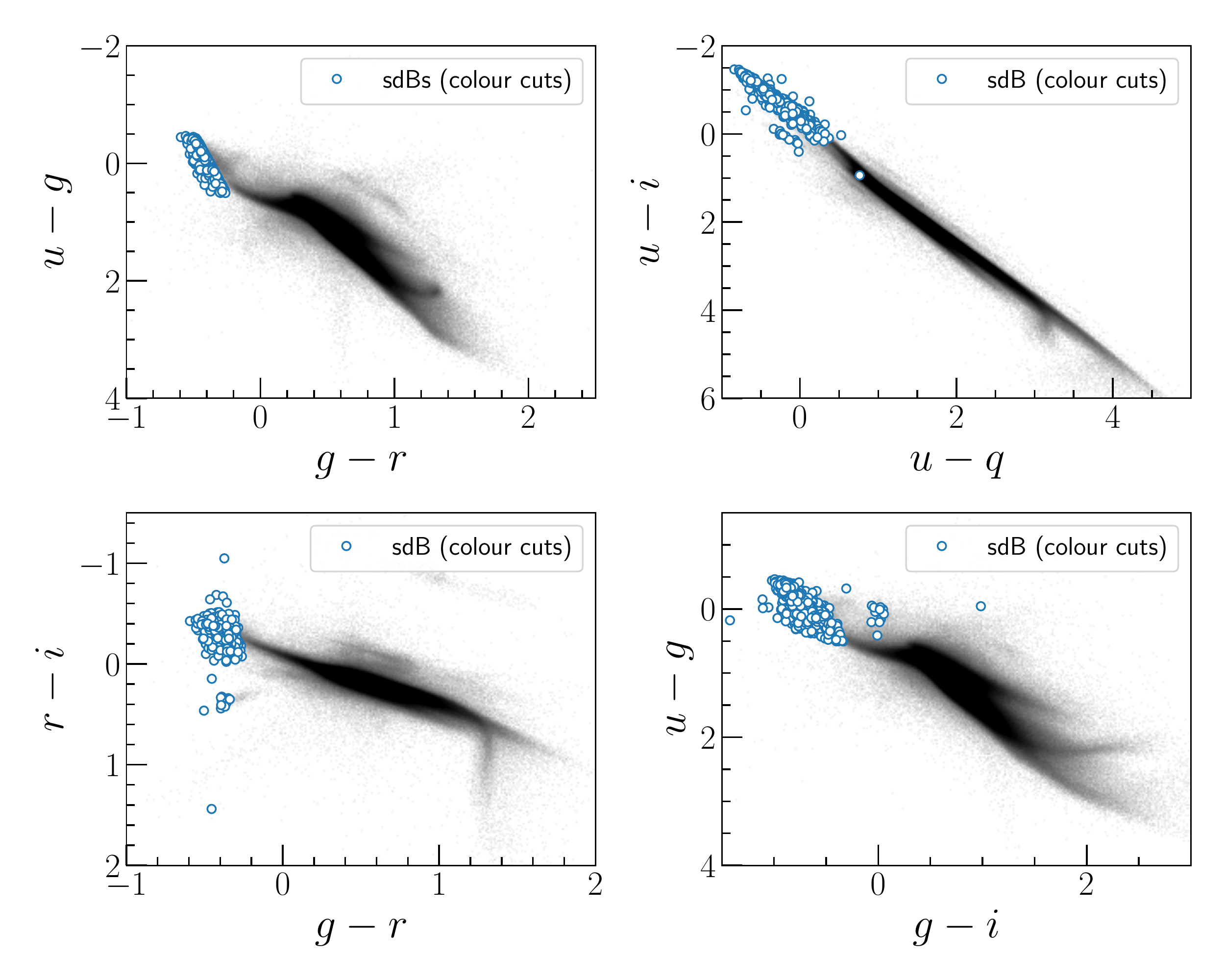}
  \caption{CCDs using different MeerLICHT colours. The sdB candidates shown in Fig.   \ref{fig:cmd_ml_gaia} are highlighted in orange here.}
  \label{fig:ccd}
\end{figure*}

Colour-colour diagrams (CCDs) help us further visualise and identify different stellar populations through their different colour properties. Figure~\ref{fig:ccd} shows four different MeerLICHT CCDs, revealing different sub-populations of stars depending on the colours used. We exploit these CCDs to identify sdB candidates in the MeerLICHT data using the colour classification schemes developed by \cite{Geier2020} for the SDSS filter set. \citet{Geier2020} define the class of hot subdwarfs by SDSS colours as follows:\\
\begin{align*}
&&0.5 < (u - g)_{\rm SDSS} < 0.7, ~~~~~~~~~~~~~~~~~~~~~~~~~\\
&&(g - r)_{\rm SDSS} > 0.208\,(u-g)_{\rm SDSS} - 0.516,\\
&&(g - r)_{\rm SDSS} < 0.208\,(u-g)_{\rm SDSS} - 0.376.
\end{align*}
To apply these colour selection criteria to our data, we first transform the MeerLICHT colours (denoted as ML) to SDSS colours using the following conversions, derived from areas where there is overlap between the two surveys: \\
\begin{align*}
&(u-g)_{\rm SDSS} = 1.1529\,(u-g)_{\rm ML} + 0.1174,\\
&(g-r)_{\rm SDSS} = (g-r)_{\rm ML} + 0.0224,\\
&g_{\rm ML} = g_{\rm SDSS} - (0.0109 \pm 0.0017),\\
&r_{\rm ML} = r_{\rm SDSS} + (0.0115 \pm 0.0022),\\
&i_{\rm ML} = i_{\rm SDSS} + (0.0101 \pm 0.0049),\\
&u_{\rm ML}=u_{\rm SDSS} - (0.1127 \pm 0.0029) \\
& ~~~~~~~~~~ - (0.1326 \pm 0.0067) \times (u-g)_{\rm SDSS}.\\
\end{align*}
Applying the above selection criteria yields 2\,188 sdB candidates within the cross-matched and quality-flag filtered MeerLICHT data. These candidates are shown in blue in Figs.~\ref{fig:cmd_ml_gaia} and \ref{fig:ccd}. We note that the selection criteria do not discriminate against white dwarf stars. Furthermore, these cuts do not account for systems that contain sdB stars and a bright MS companion whose composite colour would locate the system closer to the MS. 

\subsection{Light curves \label{sec:lightcurves}}
\begin{figure}
  \centering
  \includegraphics[width=0.95\linewidth]{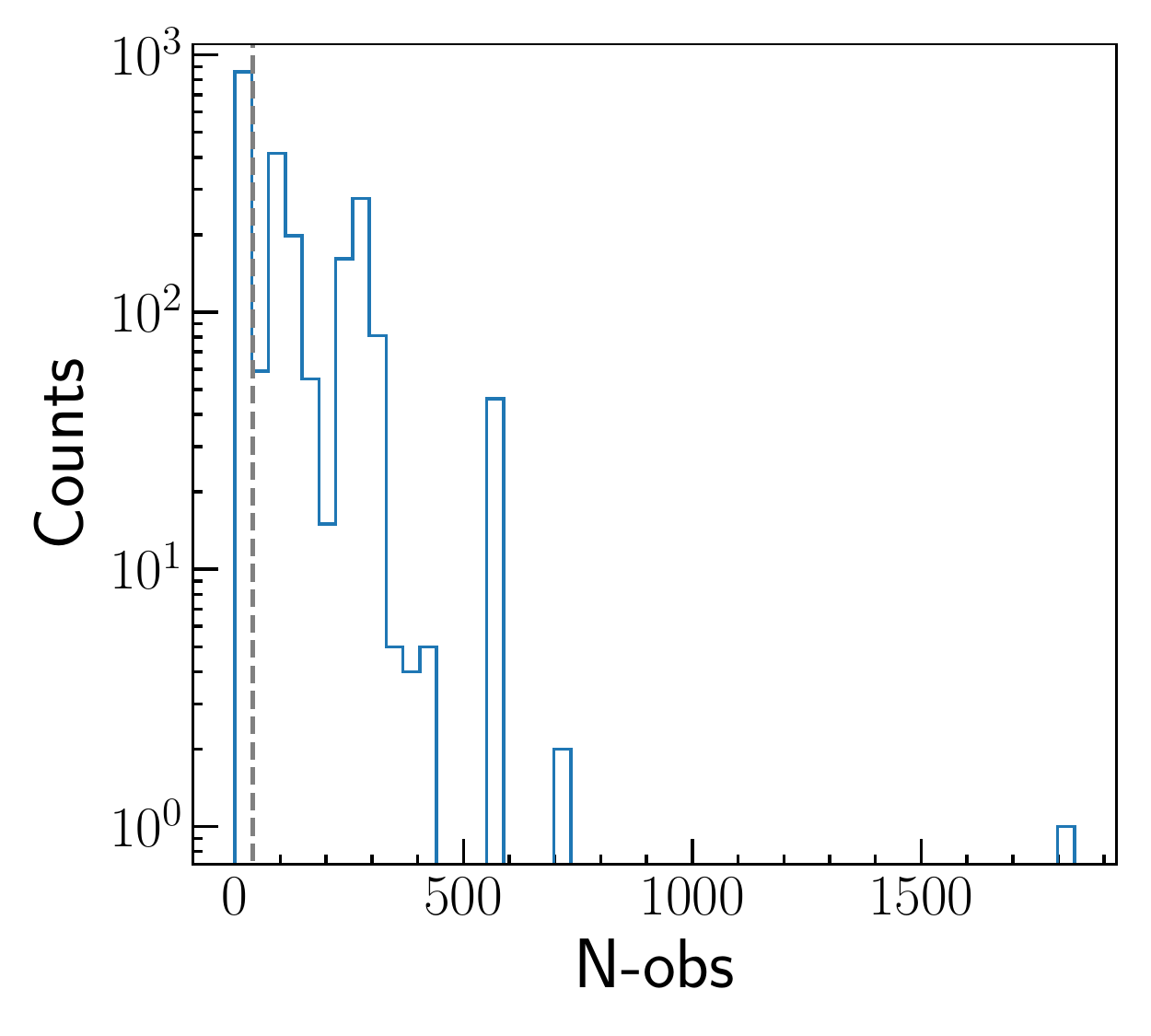}
  \caption{Distribution of the number of observations in the $q$-band filter (N-obs) for the 2188 sdB candidates. The y-axis corresponds to the number of objects per N-obs bin and is represented on a logarithmic scale. The vertical dashed grey line indicates the cutoff at N-obs = 40 in MeerLICHT $q$-band observations. All objects to the right of this cutoff line are selected to be in our sdB candidate sample.}
  \label{fig:n_hist}
\end{figure}
\begin{figure*}
  \centering
  \includegraphics[width=1\linewidth]{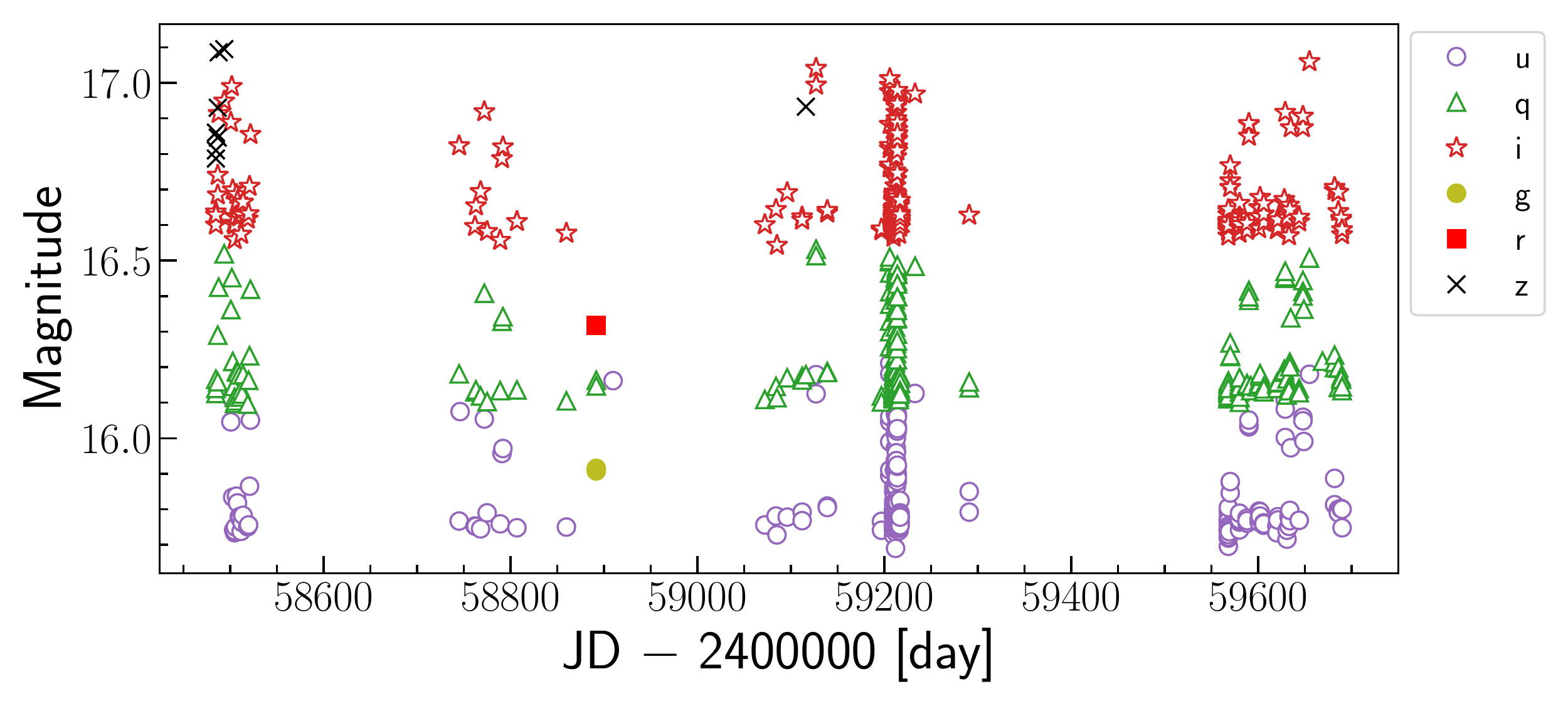}
  \caption{Illustration of the nature of MeerLCIHT's light curve for the target in the second panel of Figs.   \ref{fig:periodogram_1term} and \ref{fig:periodogram_2terms}. The irregularity in the sampling and the gapped observations can be clearly seen in this plot.}
  \label{fig:ml_lc}
\end{figure*}

We extracted light curves of the sdB candidates and analysed their time-series behaviour. This extraction was done using the forced photometry package for MeerLICHT (Vreeswijk et al., in prep). In this case, we took the coordinates of the 2\,188 sdB candidates and performed optimal photometry at the average location of the closest matching source that exists in the database. We only retained flux estimates extracted from images that do not have a `red' flag (i.e. excluding sources with large photometric zero-point uncertainties). The distribution of the number of observations in the $q$ band for each of the 2\,188 sdB candidates is shown in Fig.~\ref{fig:n_hist}. This distribution highlights the heterogeneity in the number of observations per target that is introduced by the schedule priorities. Based on Fig.~\ref{fig:n_hist}, and following the requirement adopted by \citet{DeRidder2022} in their analysis of the \textit{Gaia} DR3 time-series photometry to hunt for non-radial oscillations in main-sequence stars, we only retained sdB candidates that had at least 40 observations in the $q$ band. This limit is more conservative than that of \citet{MoralesRueda2006b}, who suggested that only 25 data points are needed to conclude that a star is variable. This left us with 610 final sdB candidates, with at least 40 MeerLICHT $q$-band measurements, to be treated by our time-series analysis methods. An example light curve for a typical sdB candidate with sparse and gapped data in multiple filters is shown in Fig.~\ref{fig:ml_lc}.

Among the 610 candidates, 44 were already identified as variables from the \textit{Gaia} DR3 catalogue \citep{Gaia2022main}. 
Our results from the frequency analysis (Sect. \ref{sec:freqmethod}) allow us to make a preliminary classification of these variables based on their phase light curve shapes. Their periodograms and phase light curves are described in Appendix \ref{appendix:A}. We summarise the characteristics of these candidates in Table \ref{tab:gaia_candidate_list}, including their photometric measurements and derived periods\footnote[1]{A full list of the candidates with their derived periods is only available in electronic form at the CDS via anonymous ftp to \rurl{cdsarc.cds.unistra.fr} (\rurl{130.79.128.5})}. In this table, we leave the variability class blank for candidates that show no evident periodic variability. The positions of the stars with a clearly variable phase curve are indicated in the CMDs in Fig.~\ref{fig:cmd_ml_gaia}. This includes some tens of eclipsing binaries (EBs; open orange circles) and a few candidate variables with sinusoidal light curves (red stars).

\begin{figure}
  \centering
  \includegraphics[width=0.95\linewidth]{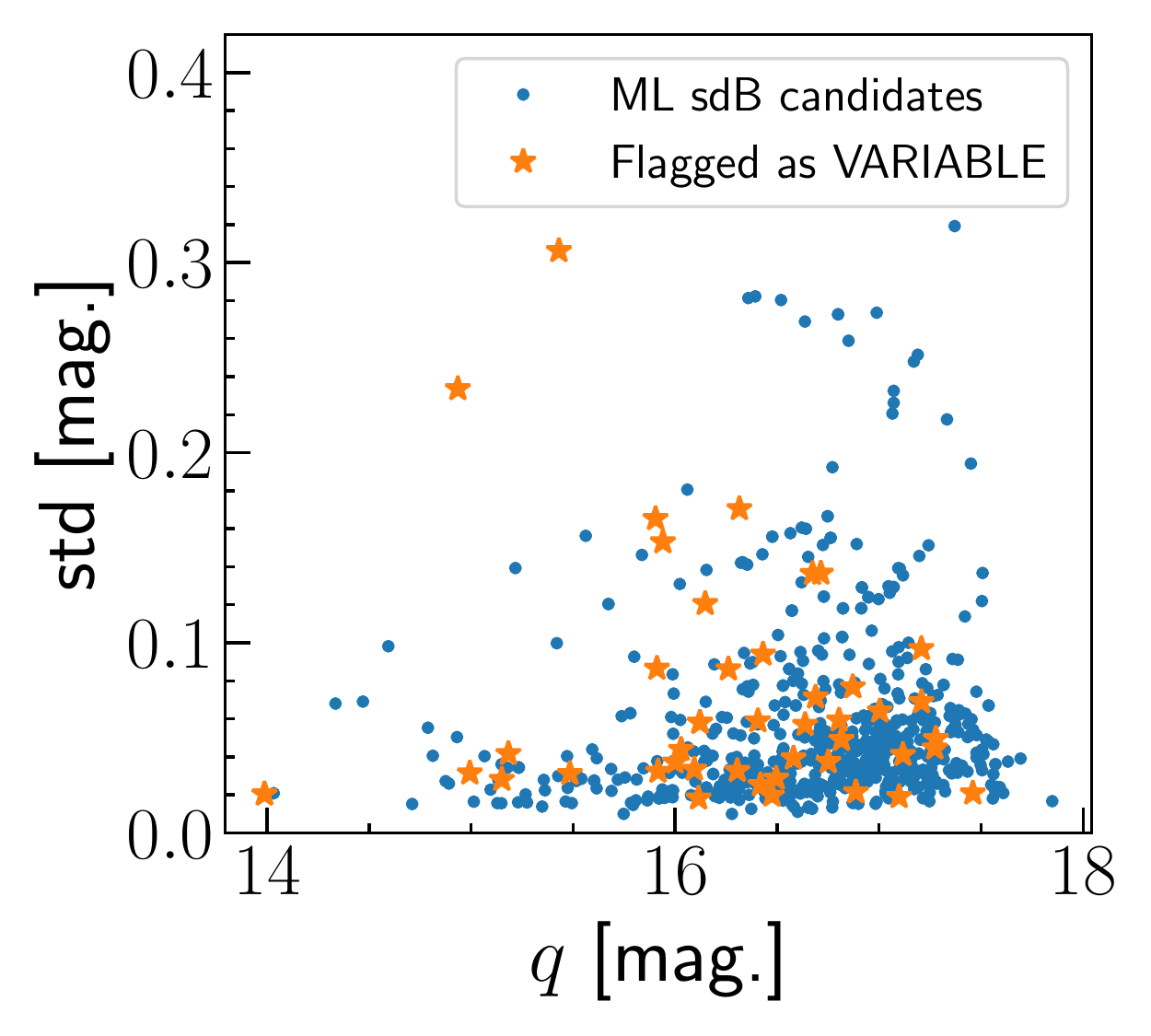}
  \caption{Magnitude in the $q$ band (qmag) versus standard deviation (std). Data points flagged as VARIABLE in the \textit{Gaia} DR3 catalogue are marked in orange.}
  \label{fig:noise_floor}
\end{figure}

With more than 40 observations in a single band, we hope to mitigate the chance that we misclassify a star as variable due to instrumental or other noise. In order to understand the behaviour of the noise in our sample, we compute the standard deviation for the $q$-band light curves for our final candidates, as shown in Fig.~\ref{fig:noise_floor}. As a comparison, we mark the targets that were flagged as {VARIABLE} in the \textit{\textit{Gaia}} catalogue in orange. We find that, generally, the light curves of the MeerLICHT sources have similar scatter compared to those of the validated \textit{Gaia} variables, indicating that the ground-based MeerLICHT data are suitable for identifying and characterising the time-series variability of compact stars.

\section{Frequency analysis of sparsely sampled multi-colour data}
\label{sec:freqmethod}

Since sdB stars are known both to exist in binaries and to host self-driven stellar oscillations, we aim to identify the dominant variability in our sdB candidates. The limitations of frequency analysis, notably the iterative process of identifying and removing dominant periodic signals (also known as pre-whitening), are determined by the quality of the astronomical dataset, for example, the time base of the data, the regularity and duty-cycle of the data, and the precision of the data. As previously discussed, MeerLICHT data was irregularly gapped and suffer from variable local weather conditions. Additionally, MeerLICHT takes data in multiple filters, which presents a challenge to traditional Fourier-based frequency analysis methods that assume homoskedastic, regularly sampled data from a single passband. 

In this section we first discuss statistical methods that have been developed for applications in the time domain (e.g. \citealt{Stellingwerf1978}, \citealt{Reimann1994}), as well as in the Fourier domain (e.g. \citealt{Lomb1976,Scargle1982}). Although these methods have been shown to work very well for dedicated asteroseismic ground-based campaigns \citep[e.g.][]{Breger1993}, it remains challenging to capture the frequency information from sparsely sampled and multi-colour light curves of faint objects such as those treated in this work. Specifically, we develop a better technique that works for the most challenging time-series data in astronomy. Our method relies on a combination of statistical time-domain and Fourier-based techniques inspired by the work of \cite{Saha2017} and \citet{VanderPlas2018}, to which we refer for details and motivation on why such a scheme is beneficial for data as treated here. In the following sections we highlight the key points of the method while focusing on the improvements made in our new implementation of this technique.

\subsection{The Lafler-Kinman $\Theta$ statistic}
The Lafler-Kinman (LK) statistic is a non-parametric method for searching for periodicity in time-series data that was originally developed to find the pulsation periods of RR Lyrae stars from single-band data \citep{Lafler1965}. Based on the underlying principle of the LK statistic, the true pulsation period P in a light curve is the one that shows minimal scatter in the phased light curve when folded with the candidate period. For a given period P, the LK statistic as formulated by \cite{Lafler1965} is given by
\begin{align}
 &  & \Theta = \frac{\sum_{i=1}^N (m_i - m_{i-1})^2}{\sum_{i=1}^N (m_i - \overline{m})^2}\ ,
\end{align}\label{eq:lk}
where $N$ corresponds to the number of observations, $m_i$ are the magnitudes at times $t_i$, and $\overline{m}$ is the mean of $m_i$. The magnitudes $m_i$ in this equation are sorted in ascending order of the time series folded with the period P as\begin{align}
& & \phi_i = (t_i\, /\rm\, P)\, modulo\, 1,    
\end{align}
where $\phi_i$ is the phase for a given $m_i$. 

Finding the correct period in a light curve is the subject of numerous studies building on the original LK statistic. One of those is
the well-known phase-dispersion minimisation (PDM) introduced by \cite{Stellingwerf1978}. With this PDM method,  the dispersion is computed by binning the phase light curve into a certain number of bins. The sum of the scatter in each bin provides the general level of periodic variability present in the light curve. The LK statistic is a limiting case of PDM, where each bin contains a minimal number of two data points. This choice is preferable for sparse datasets \citep{Saha2017}.

In case of noisy sparse observations such as in MeerLICHT, it is important to include the uncertainties of the measurements, $\sigma_i$, into the computation of the statistic, denoted here as $\Theta$. We implemented this by adding weights $w_i$ to each data point. Following \cite{Saha2017}, the weights $w_i$ and the modified LK statistic are given by
\begin{align}
    && w_i = \frac{1}{\sigma_i^2 + \sigma_{i-1}^2}\ 
\end{align}
and
\begin{align}\label{eq:theta_stat}
  &  & \Theta = \frac{\sum_{i=1}^N w_i (m_i - m_{i-1})^2}{\sum_{i=1}^N (m_i - \overline{m})^2 \, \sum_i^N w_i}.   
\end{align}
Here, $\overline{m}$ is the weighted average of $m_i$. The period that gives the smallest value of $\Theta$ is considered to be the best estimate of the correct period. However, this is not always true in the case of noisy and unevenly sampled data, because strong false alias peaks may occur in the periodogram.

\subsection{The Lomb-Scargle periodogram}
Fourier analysis is a widely used method for searching for frequency in time-series data. This approach has been shown to be effective for uniformly sampled data. Considering a continuous time series of data $x(t)$, its Fourier transform is given by
\begin{align}
    && F(\nu) = \int_{-\infty}^{+ \infty} x(t) e^{-2\pi i \nu t}\,dt\, ,
\end{align}
where $\nu$ is the cyclic frequency and $i^2=-1$ the imaginary unit. The so-called power spectral density, or PSD($\nu$), is defined as the squared amplitude of $F(\nu)$:
\begin{align}
    && {\rm PSD(\nu)} = |F(\nu)|^2\ .
\end{align}
The PSD($\nu$) contains all the frequency information in the data in the case of a perfect and continuous signal. In ground-based astronomy, observations are taken on an irregular basis  due to observational constraints such as telescope scheduling, weather, and seasonal cycles. These constraints lead to unevenly sampled time-series data with frequent large gaps.

A first approach for coping with such discrete time-series data (referred to as $x_n$), known as the Schuster periodogram (or classical periodogram), was proposed by \cite{Schuster1898}:
\begin{align}
&& P_S(\nu)= \frac{1}{N} \left| \sum_{n=1}^N x_n e^{-2 \pi i \nu t_n} \right| ^ 2\, .
\end{align}
For uniformly sampled data, the Schuster periodogram can recover all the frequency information contained in the data. In a statistical sense, this periodogram is an estimator of the PSD \citep{Scargle1982}. 

For non-uniform sampling, \cite{Scargle1982} proposed a generalisation of the classical periodogram. Following the notation in \cite{VanderPlas2018}, the Lomb-Scargle (LS) periodogram
is given by
\begin{align}\label{eq:lsp}
  P_{LS}(\nu) =
  \frac{1}{2} \Bigg\{ &
  \bigg(\sum_n x_n \cos(2\pi \nu [t_n-\tau])\bigg)^2 \bigg/
  \sum_n \cos^2(2\pi \nu [t_n-\tau]) &\nonumber\\
  & + ~ \bigg(\sum_n x_n \sin(2\pi \nu [t_n-\tau])\bigg)^2 \bigg/
  \sum_n \sin^2(2\pi \nu [t_n-\tau]) \Bigg\}\, ,
\end{align}
where $\tau$ is defined as
\begin{align}
 && \tau = \frac{1}{4\pi \nu}\tan^{-1}\Bigg(
  \frac{\sum_n \sin(4\pi \nu t_n)}{\sum_n \cos(4\pi \nu t_n)}\Bigg)\, .
\end{align} \label{tau}
This formulation was constructed in such a way that the periodogram is not dependent on the chosen zero-point for the time stamps of the data.

An interesting aspect of the LS Periodogram in Eq.\,(\ref{eq:lsp})
is its connection to a least-squares regression \citep{Lomb1976} of a harmonic model fitted to the data at a given frequency $\nu$ (see \citealt{VanderPlas2015} \& \citealt{VanderPlas2018}, for in-depth discussion). In other words, the resulting periodogram constructed from the $\chi^2$ goodness-of-fit is equivalent to the LS  periodogram. Based on this connection, \cite{VanderPlas2015} developed a generalised and multi-band periodogram derived from  minimising the $\chi^2$ of a sinusoidal model $y_k(t|\omega,\theta)$ with $\omega=2\pi\nu$ the angular frequency and $\theta$ the model parameters. This periodogram is given by
\begin{align}\label{eq:chi_min}
    && \chi^2_{min} = \chi_0^2 [1 - P_N(\nu)]\ ,
\end{align}
where $P_N(\nu)$ is a normalised version of the periodogram in Eq.\,(\ref{eq:lsp}). In this expression, $\chi_0^2$ is called the reference model, with $\chi_0^2=\sum_k(y_k/\sigma_k)$, where $\sigma_k$ denotes the error associated with the observation $y_k$. The sinusoidal model $y_k(t\,|\,\omega,\theta)$, $\chi^2_{min}$, as well as the other parameters are described in \cite{VanderPlas2015}. In their work, the authors also provide a matrix formulation of the periodogram for easy implementation.

This formulation can further be generalised to consider data of the same object taken in different photometric bands, either simultaneously or contemporaneously. In the generalised form, $y_k(t\,|\,\omega,\theta)$ is a combination of: \textit{(i)} $N_{base}$ number of Fourier components, known as the base model, which treats and fits the data as a single time series by ignoring filter labels; \textit{(ii)} a set of $N_{band}$ Fourier components, which fits each filter independently \citep{VanderPlas2015}. This combination ensures that the model does not under-fit (if using only a single standard LS model) or over-fit the data (if we only compute a standard LS periodogram for each filter and use their combined $\chi^2$ to obtain the multi-band periodogram). For each filter $k$, $y_k(t\,|\,\omega,\theta)$ is expressed as
\begin{align}
  \label{eq:xband_model}
  y_k(t|\omega,\theta) = & \ \theta_0 + \sum_{n=1}^{N_{base}} \left[\theta_{2n - 1}\sin(n\omega t) + \theta_{2n}\cos(n\omega t)\right]\ + \nonumber\\ 
  &\theta^{(k)}_0 + \sum_{n=1}^{N_{band}} \left[\theta^{(k)}_{2n - 1}\sin(n\omega t) + \theta^{(k)}_{2n}\cos(n\omega t)\right],&
\end{align}
where, $\theta_0$ and $\theta^{(k)}_0$ denote the model offsets, $N_{base}$ is the number of sinusoidal components of the base model, and $N_{band}$ the number of sinusoidal components that fit the data for each filter $k$. 

\subsection{A hybrid method: The $\Psi$ statistic}
\begin{figure*}
  \centering
  \begin{tabular}{c}
      \includegraphics[width=0.9\linewidth]{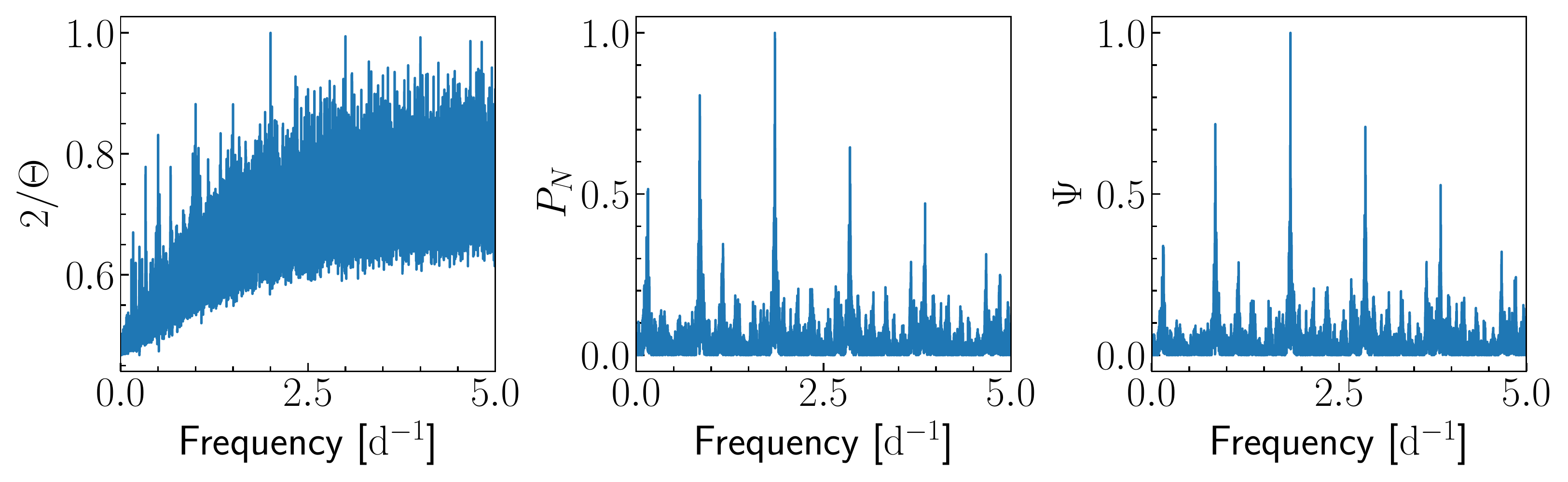}\\
      \includegraphics[width=0.9\linewidth]{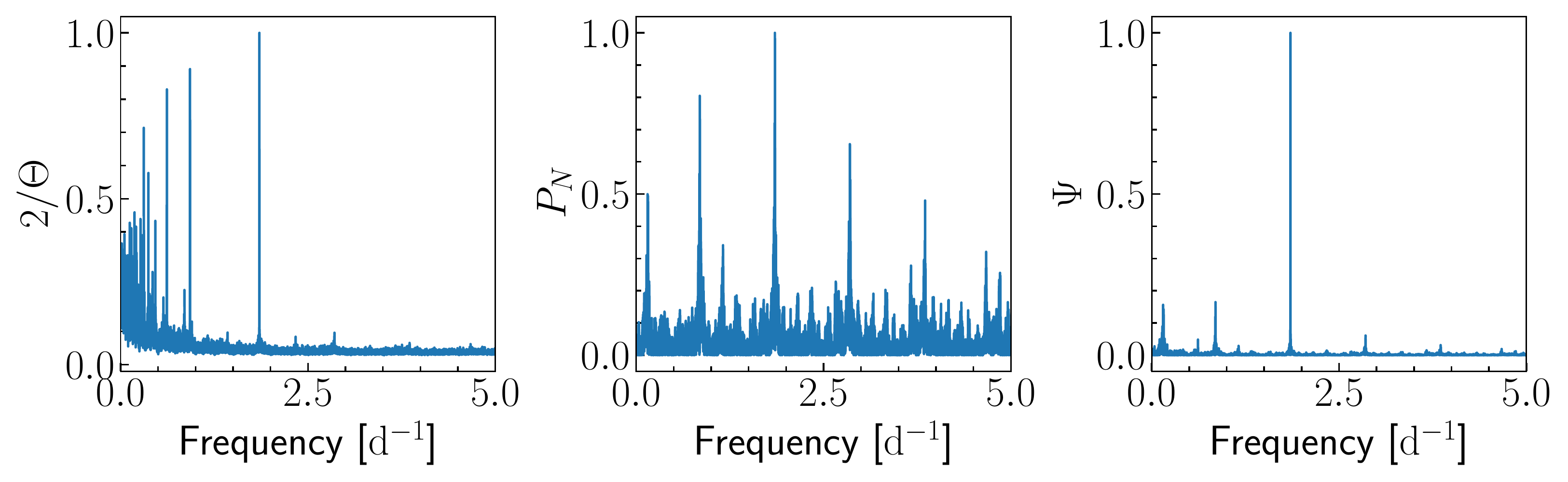}
  \end{tabular}
\caption{Illustration of the difference between scaling the magnitudes to have the same amplitude before computing the periodogram (bottom panel) and without scaling the magnitudes (top panel). The improvement in the cleanliness of the periodogram using $\Psi$ is shown in the right panels. These plots correspond to the second target in Table  \ref{tab:periods_2terms} or a zoomed-in version of the top middle panel in Fig.~\ref{fig:periodogram_1term}}.
  \label{fig:periodogram_ml_comp}
\end{figure*}

Combining the time-domain and Fourier-based methods is efficient for searching frequencies in time-series data that present an irregular (i.e. non-sinusoidal) shape and are sparsely sampled. For this reason \citet{Saha2017} introduced the so-called $\Psi$ statistics, defined as
\begin{align}\label{eq:psi}
    &&\Psi=2\Theta/P_N\ ,
\end{align}
where $\Theta$ and $P_N$ are derived from 
Eqs.~\ref{eq:theta_stat} and \ref{eq:chi_min}, respectively. 
Normalised periodograms have been used before in the exploitation of combined sparse datasets of the same variable star \citep[e.g.][]{Aerts2006,Aerts2017} but this was done without relying on a firm mathematical basis. With our introduction of the normalised periodogram into the framework as in Eq.~\ref{eq:psi} we improve upon the use of such a quantity in applications to multiple time series of a star. Since the window function behaves differently in the LK statistic and the normalised periodogram, the combined $\Psi$ statistic effectively suppresses aliased frequencies, resulting in a cleaner periodogram compared to each of $\Theta$ and $P_N$ separately, as illustrated in the columns of Fig.~\ref{fig:periodogram_ml_comp}. In summary, we made use of the following new implementations to perform frequency analysis.

First, we used the so-called generalised multi-band periodogram developed by \cite{VanderPlas2015} and adopted an implementation with a Fourier-based component as a hybrid period search method. This generalised method is optimised to take the sparse and irregular nature of our observations into account  and, most importantly, allows for an easy exploration of the multi-band filters we have. It also includes the uncertainties in the flux or magnitude measurements in the computation of the periodogram, which is crucial in the case of noisy observations.

Second, the frequency grid search is defined in such a way that our algorithm can search for high-frequency variability.

Finally, with the goal of improving the quality of the periodogram, we scaled the magnitude values such that each filter has the same amplitude before we computed the periodogram.\\

Following our tests and science goals, our periodicity search was conducted using the scaled composite light curve for both the LK statistic and the LS periodogram, and setting $N_{base}=1$ and $N_{band}=0$ for the LS periodogram (i.e. we ignore the two terms $\theta^{(k)}_0$ and $N_{band}$ in Eq. \ref{eq:xband_model}). Since MeerLICHT has narrower time spacing (time difference between two consecutive observations), particularly in the composite light curves, we can search for short-period variability in the sdOB candidates.

\subsection{Defining the frequency grid}

After defining the two elements ($\Theta$ and $\Pi$) of the $\Psi$ statistic, we must now define the trial frequencies at which we want to search for periodicities in the data. Establishing a universal method for defining this frequency range is not trivial, particularly for irregularly sampled data. Therefore, it is crucial to choose the optimal frequency grid and the maximum frequency (Nyquist frequency) such that we: $(i$) optimally and robustly probe the highest frequency regimes, $(ii$) have a small enough frequency step as to not miss any periodic signals, and $(iii$) minimise the computational cost. 

In this work, and following Chapter\,5 in \citet{Aerts2010}, we adopted the minimum frequency as the inverse of the total time base ($t_{\rm base}=t_{max} - t_{min}$). In practice, since our data are unevenly sampled and exhibit some wide gaps across the time base, we multiplied this value by a factor of 4. This implies that we searched for a frequency of a signal that repeats at least four times within the time base. The frequency step is defined as the inverse of the time base, divided by an oversampling factor (10 in our case). Lastly, the maximum frequency is determined based on the maximum sampling rate MeerLICHT can achieve in one day. It takes the integration time (60 seconds) and the overhead time (30 seconds) into account. The above parameters are defined as follows: 

\begin{align}
&~~~~~~~~f_{\rm min}=4/t_{\rm base},\nonumber\\
&~~~~~~~~f_{\rm max} = \frac{86400}{2\,(t_{\rm overhead} +t_{\rm integration})},\\
&~~~~~~~~\Delta f = \frac{1}{(10 \cdot t_{\rm base})}, \nonumber\\
\nonumber
\end{align}
where the value 86400 corresponds to the number of seconds in one day, $\Delta f$ the frequency step, and $t_{\rm overhead}$ and $t_{\rm integration}$ are the overhead and integration times, respectively. Both $f_{\rm min}$ and $f_{\rm max}$ are expressed in the number of cycles per day ($\rm d^{-1}$). We note that using this definition of $f_{\rm max}$ gives us a maximum frequency of 480 ($\rm d^{-1}$). 

We used slightly different frequency grid settings for the \textit{Gaia} data since it has different observational characteristics: as described in \cite{Aerts2010}, the minimum frequency is kept as the inverse of the time base ($f_{\rm min}=1/t_{\rm base}$); the maximum frequency as half the inverse of the median of the difference of two consecutive observations ($f_{\rm max}=1/(2 \times median(t_{i+1}-t_i))$); and the frequency step as the ratio between $f_{\rm min}$ and an oversampling factor of 10 ($\Delta f = f_{\rm min}/10$).

\subsection{Multi-colour versus composite light curves}
Since we have data in multiple passbands, we test how to best incorporate the multiple data streams into our periodicity search. The generalised LS periodogram allows for the direct inclusion of data in multiple passbands via the $N_{band}$ term. However, it has been shown that adding more Fourier terms to the model increases its complexity and the background noise of the resulting LS periodogram \citep{VanderPlas2015}. In particular, \citet{VanderPlas2015} demonstrate that the best results are obtained when $N_{base} > N_{band}$. As we want to fully exploit our data, we devise a scheme whereby we scale the light curves of each filter to have the same root-mean-square (RMS) scatter as the $q$-band light curve, then merge all of the light curves into a single, scaled composite light curve, sorted by ascending time. We use the single scaled composite light curve to calculate both the LK statistic and the LS periodogram. We compare the resulting $\Theta$-statistic, LS, and $\Psi$-statistic periodograms using the multiple light curves and the generalised LS periodogram and the scaled composite single light curve in the top and bottom panels of Fig.~\ref{fig:periodogram_ml_comp}. We note that there is a clear improvement in the resulting $\Theta$-statistic periodogram when using the single scaled composite light curve, which in turn produces a $\Psi$-statistic periodogram with a lower noise level.

Our RMS scaling scheme is not without consequence, however. The light curve scaling is not perfect, and therefore results in a small, but still present amplitude offset between data points in different filters. When paired with MeerKAT or while on the backup fields, MeerLICHT observes in a repeated $uqi$ filter sequence. The exposure time, readout, and filter changing result in a repeated measurement in a given filter roughly every five minutes. This introduces a five-minute signal into the window function of targets with observations following this sequence. This window pattern can be seen in periodograms for the majority of our targets, including the remaining figures shown in this work.
\subsection{Window function pattern}
\begin{figure}
  \centering
  \includegraphics[width=0.95\linewidth]{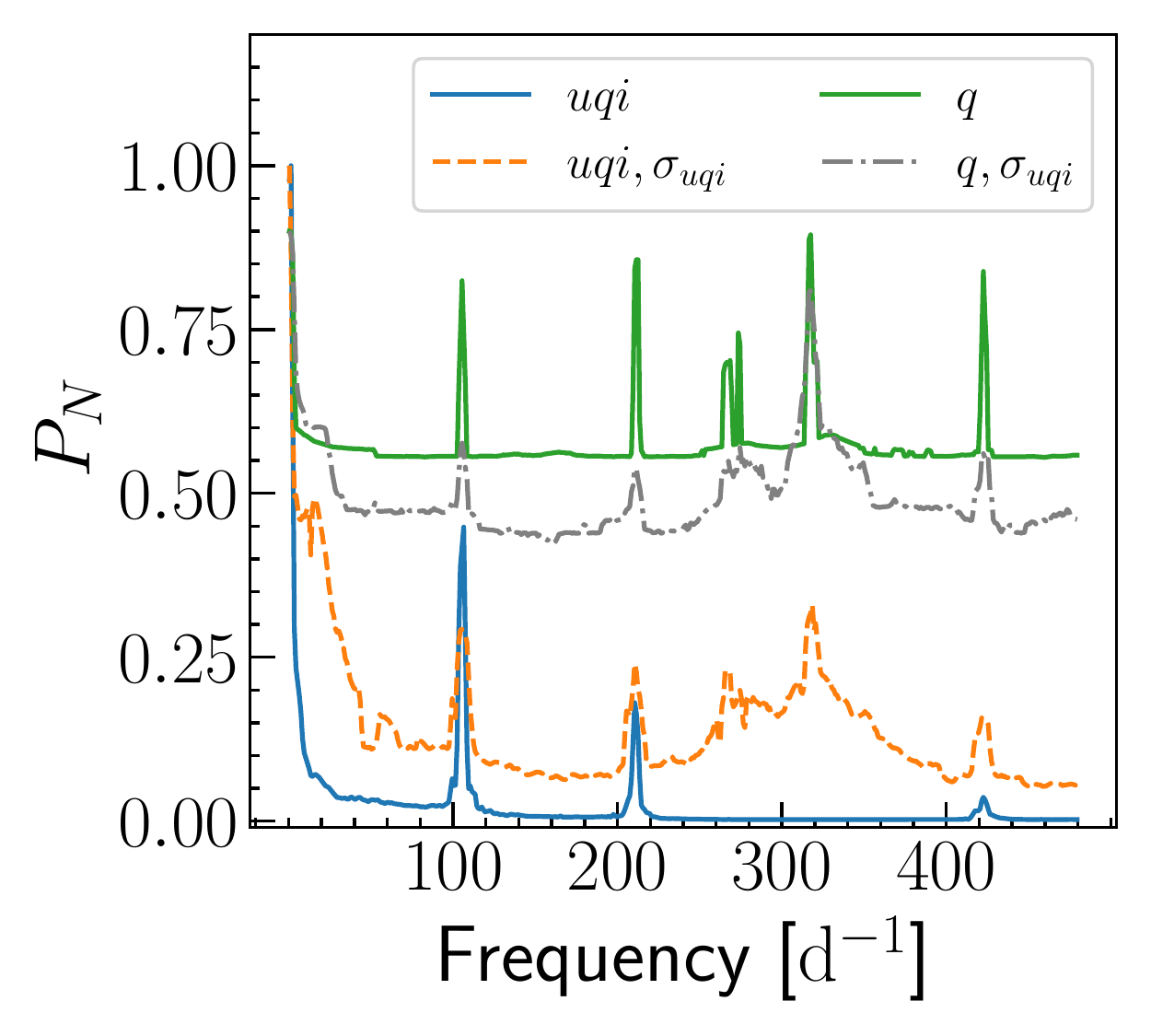}
  \caption{Averaged LS periodograms of the candidates' window functions using a frequency bin width of 1 cycle per day. The solid blue and green lines correspond to the window functions in the composite $uqi$-band filters and the $q$ band, respectively. The dashed orange and grey lines represent the same window functions, but with the contribution of the uncertainties of the magnitude values included in the calculation of the periodograms.}
  \label{fig:hist_wfunc}
\end{figure}

It is crucial to study the window function to better understand different patterns occurring in the resulting periodograms. In other words, knowing the nature of the window function helps us detect aliases in our periodograms. With this in mind, we built an averaged periodogram of the candidates’ window functions, where the spectral powers, $P_N$, are averaged per frequency bin width of 1 $\rm d^{-1}$. Here, we examine the window functions of the composite $uqi$-band light curves and those of the $q$ band alone (in which objects are most observed), such that we can investigate the effects of combining the light curves on the window function pattern.
Additionally, we included the uncertainties in the magnitudes while computing the periodograms to reveal their possible contributions to the shape of the window function. The resulting periodograms are presented in Fig. \ref{fig:hist_wfunc}, where the blue and green solid lines are obtained from the periodograms of the standard window functions, whereas the orange and grey dashed lines are obtained from the introduction of the magnitude uncertainties in the calculation of the periodograms. These periodograms reveal several interesting frequency peaks:

{\it Peaks at 1, 100, and 200 $\rm d^{-1}$}: Both the combined $uqi$ bands and the $q$-band window functions peak at $\sim 1$ $\rm d^{-1}$, which is the result of daily observation sequences; and at $\sim$ 100 and 200 $\rm d^{-1}$, which are probably the harmonics of the 1 $\rm d^{-1}$ peak. \\

{\it Peaks at 280 and 300 $\rm d^{-1}$}: The frequency peak at $\sim 280$ $\rm d^{-1}$, particularly for the $q$ band, corresponds to a period of about 5 minutes, could be explained by the consecutive measurements in a given filter discussed in the previous section. This peak is more evident when the $q$-band filter alone is considered. As the time resolution increases when the three filters are used, this peak is suppressed. However, this peak interestingly appears in the $uqi$ window function, along with the one at 300 $\rm d^{-1}$, when the uncertainties are introduced in the periodograms. Moreover, the pattern of this window function (a somewhat wide hump around 300 $\rm d^{-1}$) appears more often in the periodograms of our candidates (see Appendix \ref{appendix:A} for more illustrations). \\

{\it Peak at 400 $\rm d^{-1}$}: This peak appears in all of the four cases and is most probably an alias of the 100 $\rm d^{-1}$ peak. \\

In general, combination frequencies occur due to the integration and readout times, the effects of which on the window function depend on whether a composite light curve is used. Since the mentioned frequency peaks represent the general pattern of all candidates' window functions, each candidate may have a slightly different pattern. It is therefore necessary to examine each candidate's window function to explain the source of the large bump around 300 $\rm d^{-1}$ and its correlation to the magnitude uncertainties (which is not covered in this paper). Overall, these uncertainties increase the noise levels of the periodograms at all frequencies, as shown in Fig. \ref{fig:hist_wfunc}. Regardless of the magnitude uncertainties, increasing the data sampling will alleviate the aliases caused by the window functions. This is demonstrated by the $uqi$ composite light curve (blue line) and the $q$-band window functions (green line) in Fig. \ref{fig:hist_wfunc}, where we have three large peaks in the $uqi$ periodogram compared to six in the $q$ band.\\

\begin{figure}
  \centering
  \includegraphics[width=0.95\linewidth]{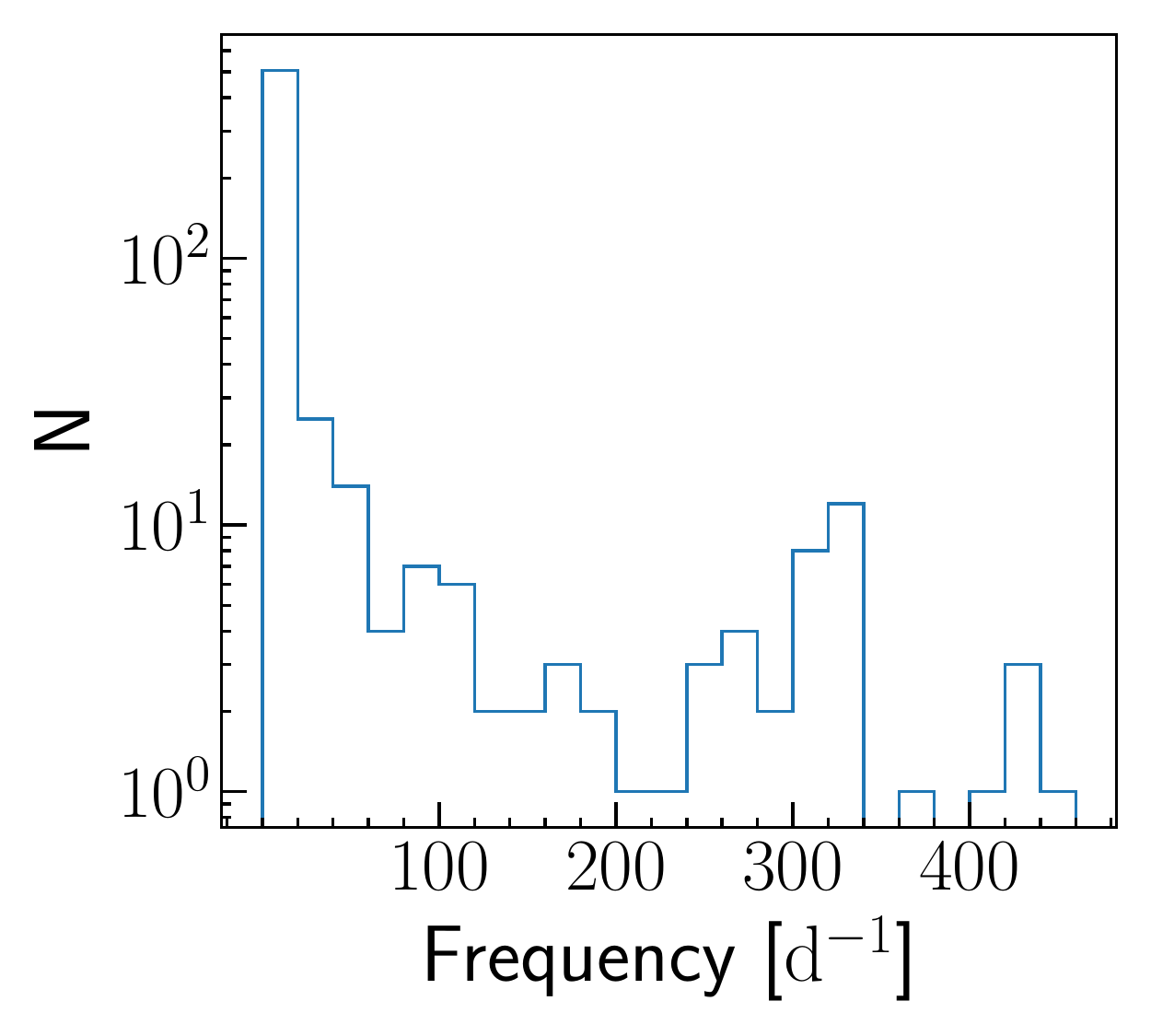}
  \caption{Histogram of the candidates' dominant frequencies found by the $\Psi$ statistic. The y-axis is on a logarithmic scale.}
  \label{fig:hist_freq}
\end{figure}

Furthermore, in Fig. \ref{fig:hist_freq}, we build a histogram of the dominant frequencies found by our algorithm for the 610 candidates. Some of the frequencies around 300 $\rm d^{-1}$ are most probably aliases due to the window function pattern in Fig. \ref{fig:hist_wfunc} (the orange dashed line). 

\subsection{Influence of the number of Fourier terms}
While PDM methods are well suited for detecting periodicities of non-sinusoidal signals, Fourier-based methods such as the LS periodogram are not. Typically, such Fourier-based methods require multiple harmonic terms to represent non-sinusoidal signals, like those produced by EBs. It is then a non-trivial task to select the base period from the resulting series of harmonics without direct intervention. 

\begin{figure*}
  \centering
  \includegraphics[width=0.95\linewidth]{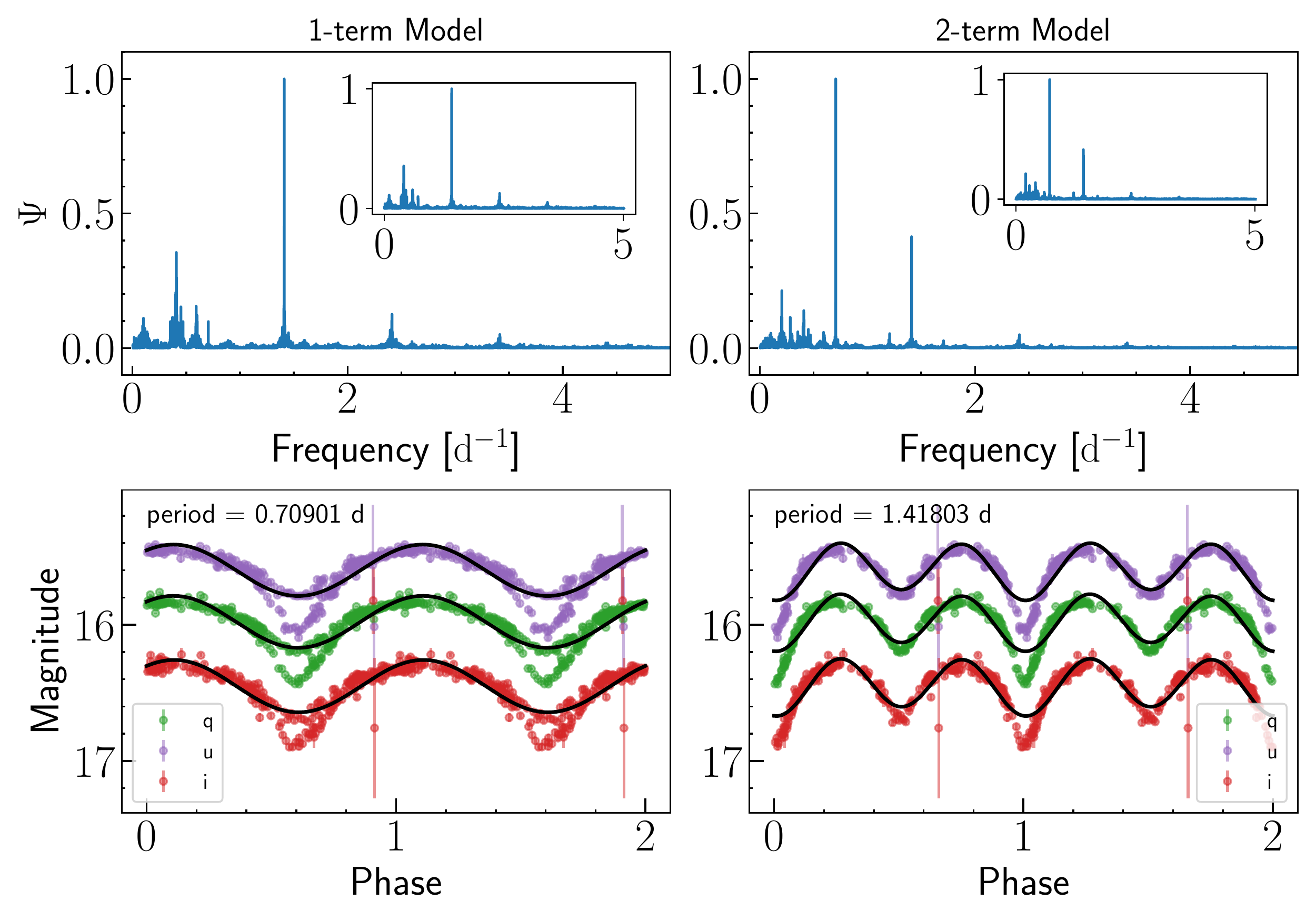}
  \caption{Periodograms and phase diagrams comparing one-term (left panel) and two-term (right panel) models. The black curve represents the model fit to the data using the dominant frequency found using the LS periodogram.}
  \label{fig:12term_model}
\end{figure*}

The generalised LS periodogram presented by \citet{VanderPlas2018} enables the inclusion of multiple harmonic terms in addition to the single base sinusoid with the term $N_{base}$. The inclusion of additional Fourier terms increases the likelihood that the true period of a non-sinusoidal signal is identified as the most dominant signal without manual intervention. An example of this is demonstrated in Fig.~\ref{fig:12term_model}, with the $N_{base}=1$ model on the left and the $N_{base}=2$ model on the right. We note that the periodograms for this example are calculated using the scaled composite light curve. In this example, we see how the $N_{base}=2$ model easily identifies the true period of the EB, whereas the $N_{base}=1$ incorrectly identifies the first harmonic (2$f_{\rm orb}$=$1/2~P_{\rm orb}$) as the dominant periodicity. However, it has been demonstrated that the inclusion of additional Fourier terms generally results in the production of a noisier periodogram \citep{VanderPlas2018}. In terms of goodness of fit, in Fig.~\ref{fig:12term_model} for instance, the two-term model slightly improves the root-mean-square error (RMSE) values in the $q$ and $u$ bands (or $\rm RMSE_q$ and $\rm RMSE_u$, respectively), with $\rm RMSE_q=0.076$ and $\rm RMSE_u=0.086$, compared to the one-term model with $\rm RMSE_q=0.082$ and $\rm RMSE_u=0.094$. 

\begin{figure*}
  \centering
\includegraphics[width=1\linewidth]{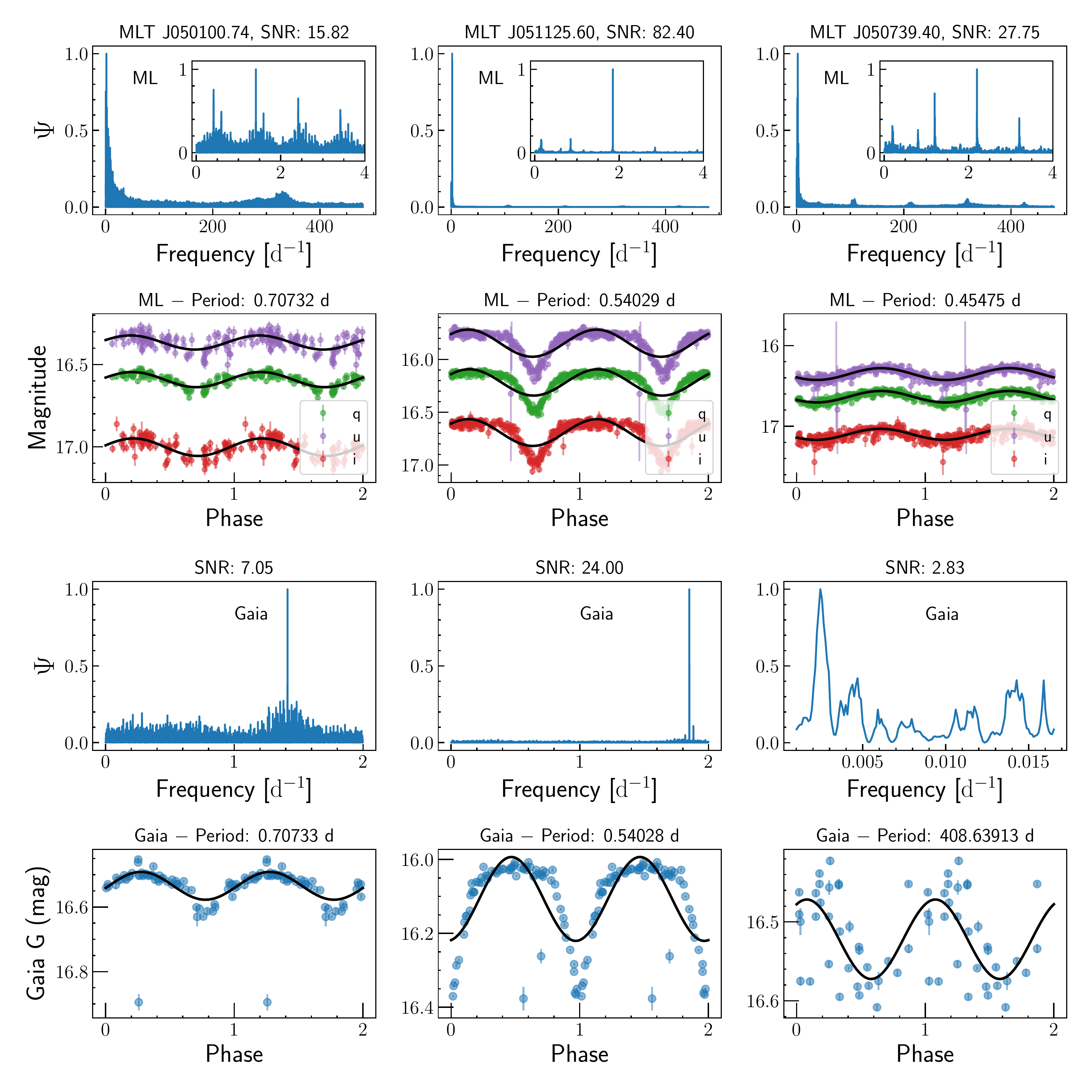}
\caption{Comparison of MeerLICHT and \textit{Gaia} periodograms and phase-folded light curves for $N_{base}=1$. Each column represents one of the three sdB candidates. The first and second rows represent the periodograms and phase-folded light curves for MeerLICHT data, respectively, whereas the last two rows correspond to that of \textit{Gaia} data. The black curve is the model fit to the data using the LS model. For illustration purposes, the phase is plotted twice.}\label{fig:periodogram_1term}
\end{figure*}

\begin{figure*}
  \centering
\includegraphics[width=1\linewidth]{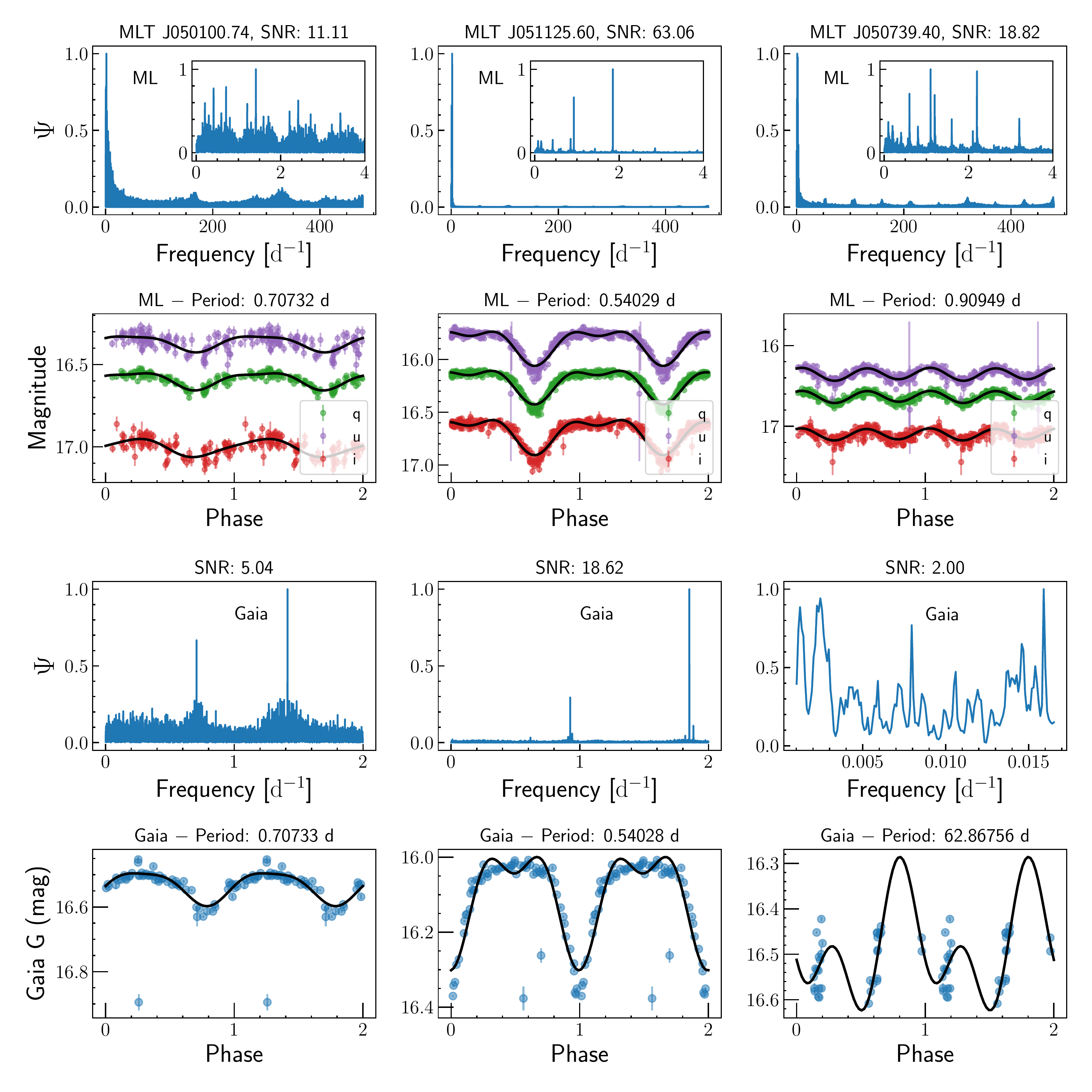}
\caption{Same as Fig.~\ref{fig:periodogram_1term}, but using $N_{base}=2$.}\label{fig:periodogram_2terms}
\end{figure*}

We further test the $N_{base}=1,2$ term models on three known variables that were observed by MeerLICHT and \textit{Gaia}, and have been previously studied by \citet{Graczyk2011}. The resulting periodograms for the $N_{base}=1$ term model are shown in Fig.~\ref{fig:periodogram_1term} and the resulting periodograms for the $N_{base}=2$ term model are shown in Fig.~\ref{fig:periodogram_2terms}. The resulting dominant periodicities found for each object are listed in Table~\ref{tab:periods_2terms}. In two of the three cases, we find that there is no difference between the identified dominant periodicities for the $N_{base}=1$ and 2 term models in the MeerLICHT data. We also find that we recover the true period, or a harmonic thereof, in all cases when using the MeerLICHT data. The third object returns unrelated periods, likely due to the fewer data points obtained by \textit{Gaia}. 

\begin{table*}
    \centering
    \caption{Summary of the derived periods.}
    \label{tab:periods_2terms}
    \begin{tabular}{cccccccc}\toprule
ML name & RA      & Dec.       & ML 1$-$term & \textit{Gaia} 1$-$term & ML 2$-$term & \textit{Gaia} 2$-$term & Graczyk et al. 2011 \\
&(deg.)     & (deg.)      & (day)       & (day)         & (day)       & (day)         & (day)                 \\
\midrule
MLT J050100.74&75.25308 & $-$67.18032 & 0.70732   & 0.70733     & 0.70732   & 0.70733     & 1.41464             \\
MLT J051125.60&77.85666 & $-$67.17393 & 0.54029   & 0.54029     & 0.54029   & 0.54029     & 1.08056             \\
MLT J050739.40&76.91418 & $-$67.67345 & 0.45475   & 408.63913   & 0.90949   & 62.86756    & 0.90950             \\
\bottomrule \\
\end{tabular}
\footnotesize{ \textbf{Notes.} The periods  are obtained from MeerLICHT (ML) and \textit{Gaia} data using $N_{base}=1$ or 2 and compared with the work of \cite{Graczyk2011}.}
\end{table*}
As expected, using $N_{base}=2$ results in higher noise level in the periodograms. To quantify this noise, we use a method similar to that used by \cite{Breger1993}, averaging the square root of the power ($\Psi$) across all frequencies except the dominant frequency peak. It is worth noting that, in \cite{Breger1993}, this noise is computed over a frequency window width of a few cycles per day around the dominant frequency. However, most of the candidates in this work exhibit somewhat high peaks at all frequencies; therefore, we compute the signal-to-noise ratio (S/N) based on the amplitude of the highest peak over the averaged noise across all frequencies.
To illustrate the noise level difference between the one- and two-term models, we annotate each candidate S/N in Figs.~\ref{fig:periodogram_1term} and \ref{fig:periodogram_2terms}. In these figures, the one- and two-term models for the first MeerLICHT target have a S/N of 15.86 and 11.09, respectively.

In their analysis, \cite{VanderPlas2018} observed a similar case and interpreted the increased background noise as directly related to the flexibility of the model in fitting the data. This explains the increase in the amplitude of the periodogram at any frequencies, not only at the true frequency, as shown in Figs.~\ref{fig:periodogram_2terms}. \cite{VanderPlas2018} further pointed out that, in some cases, the background noise could also be harmonics, which we expect to occur at $f_1/2$ for $N_{base}=2$, and at $f_1/2$ and $f_1/3$ for $N_{base}=3$, where $f_1$ is the dominant frequency. 

While the model with $N_{base}=2$ has greater flexibility in identifying the true period of EB targets, it shows little benefit in the case of sinusoidal signals such as those produced by stellar pulsations. Furthermore, including more terms greatly increases the computation time. For instance, a one-term model ($\sim$ 11 min) is about 1.45 times faster than a two-term model ($\sim 16$ min) for $\sim 6$ million trial frequencies. Considering that we are searching for general periodicities including both eclipses and pulsations, combined with the computation time and generally increased noise level in the periodogram when using $N_{base}=2$, we decide on using $N_{base}=1$ for the remainder of this work. A follow-up paper will be devoted to the derivation of the full frequency content of the light curves. Here, we focus on the dominant frequency.

\section{Sample variability characteristics}\label{sec:freq_application}
In addition to identifying the dominant period of variability, we also characterise the time-series variability of each object (i.e. the kurtosis, skewness, magnitude of variability, etc.). We calculate these variability indices using the composite light curve, and record the scaling factor between the $u$ and $i$ bands to better understand the properties of our sample of sdOB candidates.\\ 

Gaining prior knowledge of the nature of the variability of our candidates before proceeding to the frequency analysis is helpful as we can focus on the most promising candidates. To characterise the variability of the candidates, we compute diverse variability indices from their time series. For several reasons, however, we only focus on two indices: the {magnitude of variability} (V) and the {median absolute deviation} (MAD; \citealt{Rindskopf2010}). One reason is that our data contains a wide range of observations per light curve, from 40 to hundreds of observations, which might introduce a bias into the derived statistical properties. Another reason is that since the light curves are subject to outliers, we need a statistical tool that is less sensitive to outliers.\\

\begin{figure}
  \centering
  \includegraphics[width=0.95\linewidth]{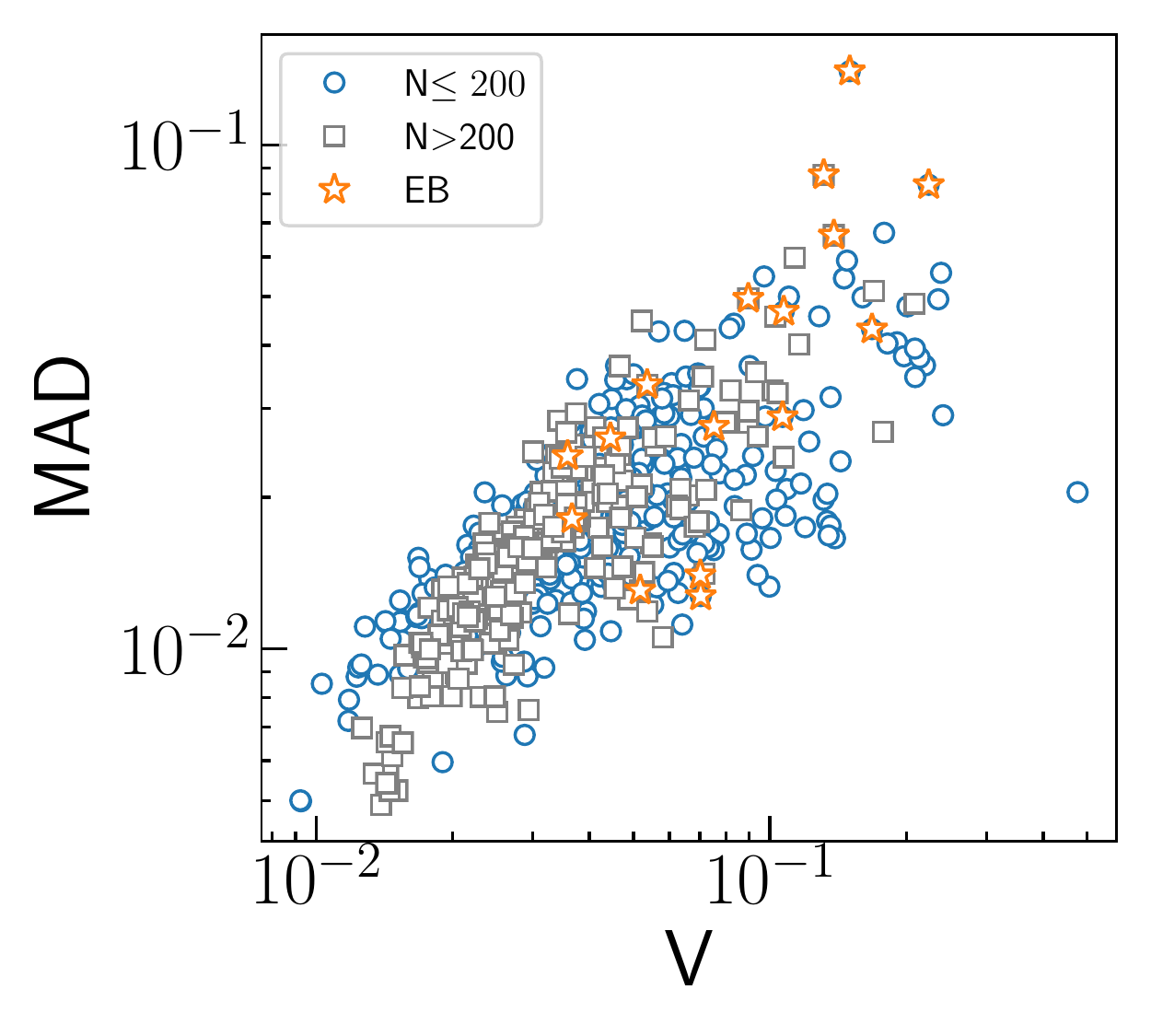}
  \caption{MAD versus the magnitude of variability index (V). The open blue circles represent objects with a number of data points N $<200$ in the $q$ band; the grey squares represent those with N$>200$. Candidates observed as EBs are represented in orange.}
  \label{fig:mad_vs_magvar}
\end{figure}

The magnitude of variability is defined as the standard deviation of the flux over the average flux. The larger this value, the higher the level of variability of a given object. On the other hand, the MAD is a robust statistical tool for measuring the variance of data, which is less sensitive to outliers compared to the standard deviation (e.g. \citealt{Eyer2022}; \citealt{Rindskopf2010}). We computed these two indices for the 610 candidates using flux values in the $q$ band. As their resulting values are somewhat skewed, we plot them on a logarithmic scale in Fig.~\ref{fig:mad_vs_magvar}. This figure shows that the two indices are independent of the number of observations and reveals that our candidates have a varied range of variability. By visually inspecting the candidates' phase-folded light curves, we plot in the same figure those candidates that are found to be EBs. The majority of these EB candidates have higher values of the MAD and V indices, which suggests that, although no evident clusters can be found in the figure, these two variability indices might provide us with constraints on the variability of the candidates. Furthermore, the resulting correlations between various variability indices are presented in Appendix \ref{appendix:B}.

\section{Conclusion}\label{sec:conclusion}
MeerLICHT observations allow us to study stellar populations in the southern sky, with approximately 5.5 million objects observed in the $u, g, r, q,$ and $i$ filters. From \textit{Gaia} DR3 parallaxes, we were able to estimate the distance to these objects as well as their absolute magnitudes. Our hot-subdwarf candidates are drawn from colour classification schemes, which yields $\sim 2000$ sdB candidates. A minimum number of observations per light curve and quality flags were applied to these candidates, producing our final set of 610 sdB candidates. The frequency search algorithms developed in this work have led to the discovery of dozens of potential EBs from these candidates, which are found at the brighter end of the EHB in the CMDs, suggesting that these objects are potentially hot massive main-sequence stars. This number would increase if we further examined ambiguous candidates. We also tested our algorithm with three known variables among our candidates, previously studied by \cite{Graczyk2011}. The extraction of the \textit{Gaia} light curves for these three variables enabled us to compare their periodograms and phase-folded light curves. In most of the cases, we find the same periods using both datasets. For MeerLICHT, different patterns in the periodograms and aliases were identified by means of the window functions. Finally, we attempted to statistically characterise our candidates using variability indices, with the aim of finding any structures in the light curves.\\

The purposes of the current study were to evaluate the capabilities of MeerLICHT in detecting faint hot subdwarfs and to develop efficient techniques for finding periodicity in unevenly sampled time-series data. This paper has shown that:
\begin{itemize}
    \item With MeerLICHT, we can effectively characterise different stellar populations in the CMDs and CCDs, notably the hot subdwarf population. A comparison of the MeerLICHT CMD with that of \textit{Gaia} reveals significant similarities in terms of how the two CMDs represent various stages of stellar evolution.
    \item MeerLICHT and \textit{Gaia} have a comparable number of observations  for objects with magnitudes between 14 and 19 mag and RUWE $\lesssim 1$. In other words, under these conditions, it is most likely that there will be many matches between the two sorts of observations.
    
    \item The hybrid implementation of both statistical and Fourier-based methods coupled with the amplitude scaling of the composite light curves are effective approaches for finding the dominant frequency.
    \item Taking the uncertainties in the magnitudes into account when examining the window functions is important, at least for MeerLICHT data, as it reveals various aliases in the candidates' periodograms.  
    \item The MAD and the magnitude of the variability indices allow us to find the most probable variable among the candidates. 
\end{itemize}

As mentioned previously, the aim of this work was to evaluate the value of MeerLICHT data in characterising the variability properties of known sdB stars in the Southern Hemisphere. To that end, we have developed tools to investigate the variability of sdB stars and other variable stars within the MeerLICHT data, and we have demonstrated the value of contemporaneous multi-colour photometry for understanding stellar variability. It is also important to note that the uncertainty determination of the frequencies found was not fully explored in this work due to several main factors: the evaluation of this type of uncertainty is not straightforward due to the correlated and heteroscedastic nature of our data, the aliases present at different frequencies, and the irregularities of the data sampling affect the pattern of the window function and the resulting periodograms. Considering all of these constraints, further analyses are needed to properly assess the uncertainties, such as those discussed in \cite{VanBeeck2021}. These limitations imply that the significance of the frequency reported here is only based on the S/N. The higher the S/N, the more we rely on its corresponding frequency.\\

Future research will concentrate on the examination of well-chosen candidates using asteroseismology and the use of spectroscopic and time-series data from other surveys, such as TESS, to enhance the frequency analysis and validation of the candidates. These candidates might not only be restricted to the sdB candidates but also include the sdO population (e.g. by following the classification schemes for sdO stars discussed in \citealt{Geier2020}). Overall, the present work makes several noteworthy contributions to MeerLICHT observations of sdB stars and data analysis techniques that could be applied to related research, including the study and characterisation of variable stars that will be observed with the BlackGEM telescopes once it is fully operational.

\begin{acknowledgements}
The research leading to these results has received funding from the
KU\,Leuven Research Council (grant C16/18/005: PARADISE), from the Research Foundation Flanders (FWO) under grant agreements
G0A2917N (BlackGEM) and K802922N (Sabbatical leave awarded to CA), as well as from the BELgian federal Science Policy Office (BELSPO) through PRODEX grants for \textit{Gaia} data exploitation.
CA
  and CJ are grateful for the kind hospitality offered by the staff of
  the Center for Computational Astrophysics at the Flatiron Institute
  of the Simons Foundation in New York City during their work visit in the fall of
  2022.
  PJG is supported by NRF SARChI grant 111692. The MeerLICHT telescope is designed, built and operated by a consortium consisting of Radboud University, the University of Cape Town, the South African Astronomical Observatory, the University of Oxford, the University of Manchester and the University of Amsterdam, supported by NWO, NRF and the European Research Council. 
\end{acknowledgements}
\bibliographystyle{aasjournal}
\bibliography{final_version.bib}

\begin{appendix}
\section{MeerLICHT's sdB candidates flagged as {VARIABLE} in \textit{Gaia} DR3}\label{appendix:A}
\begin{figure}[h!]
  \centering
  \includegraphics[width=1\linewidth]{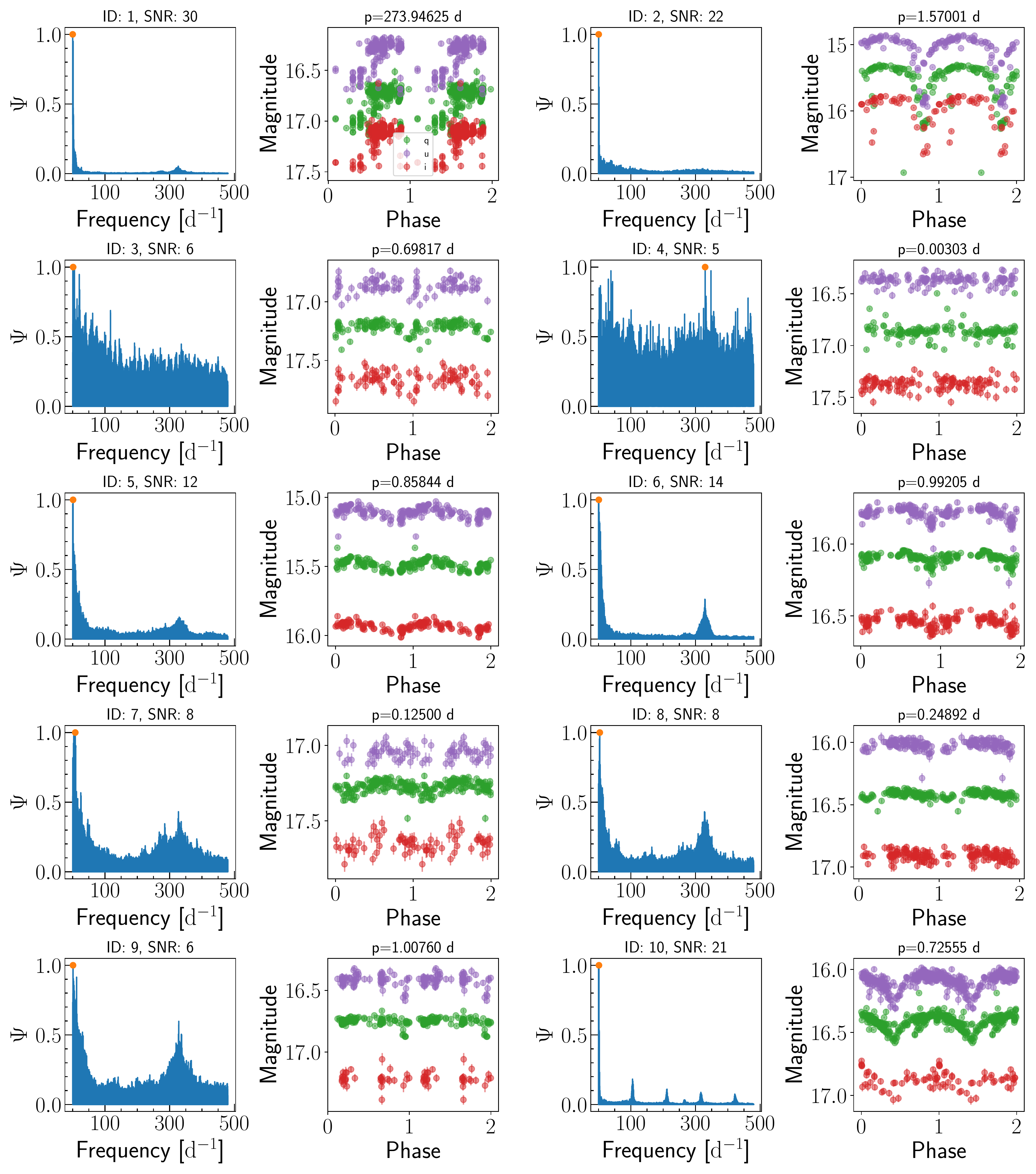}
  \caption{Periodograms and phase diagrams of MeerLICHT's candidates flagged as VARIABLE in \textit{Gaia} DR3.}
  \label{fig:gaia0}
\end{figure}

\begin{figure*}
  \centering
  \includegraphics[width=1\linewidth]{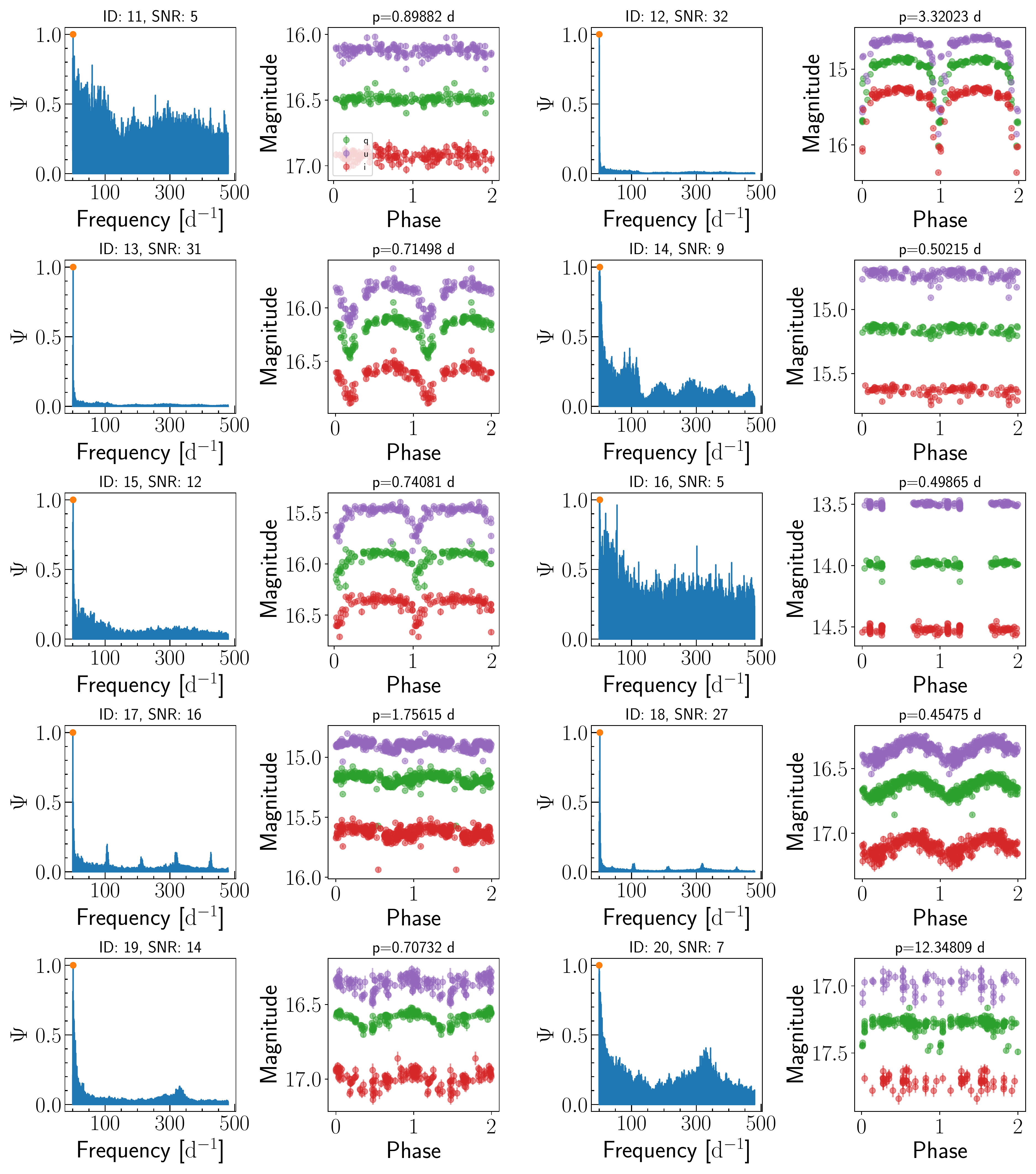}
  \caption{Periodograms and phase diagrams of MeerLICHT's candidates flagged as VARIABLE in \textit{Gaia} DR3.}
  \label{fig:gaia1}
  
\end{figure*}

\begin{figure*}
  \centering
  \includegraphics[width=1\linewidth]{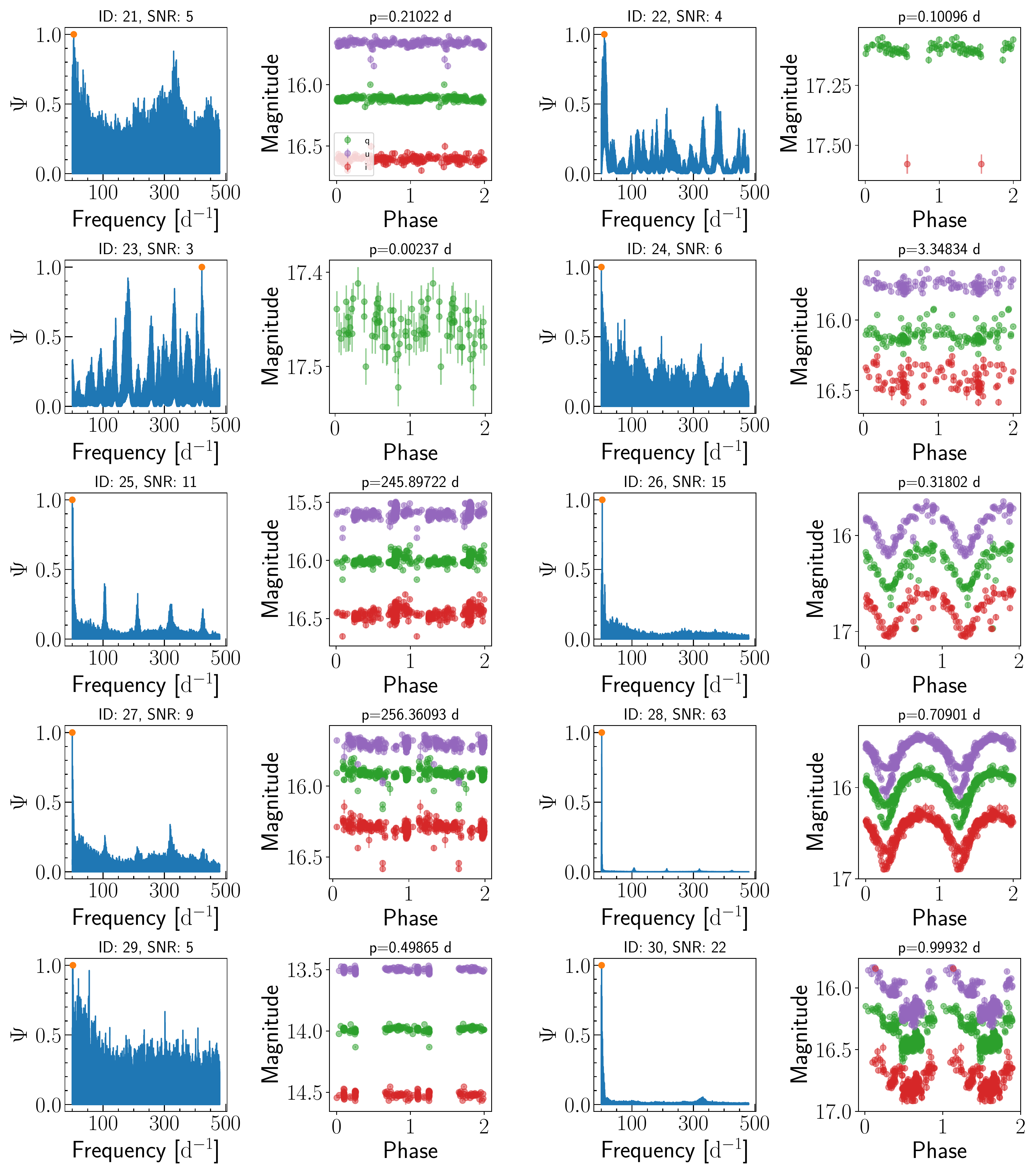}
  \caption{Periodograms and phase diagrams of MeerLICHT's candidates flagged as VARIABLE in \textit{Gaia} DR3.}
  \label{fig:gaia2}
  
\end{figure*}

\begin{figure*}
  \centering
  \includegraphics[width=1\linewidth]{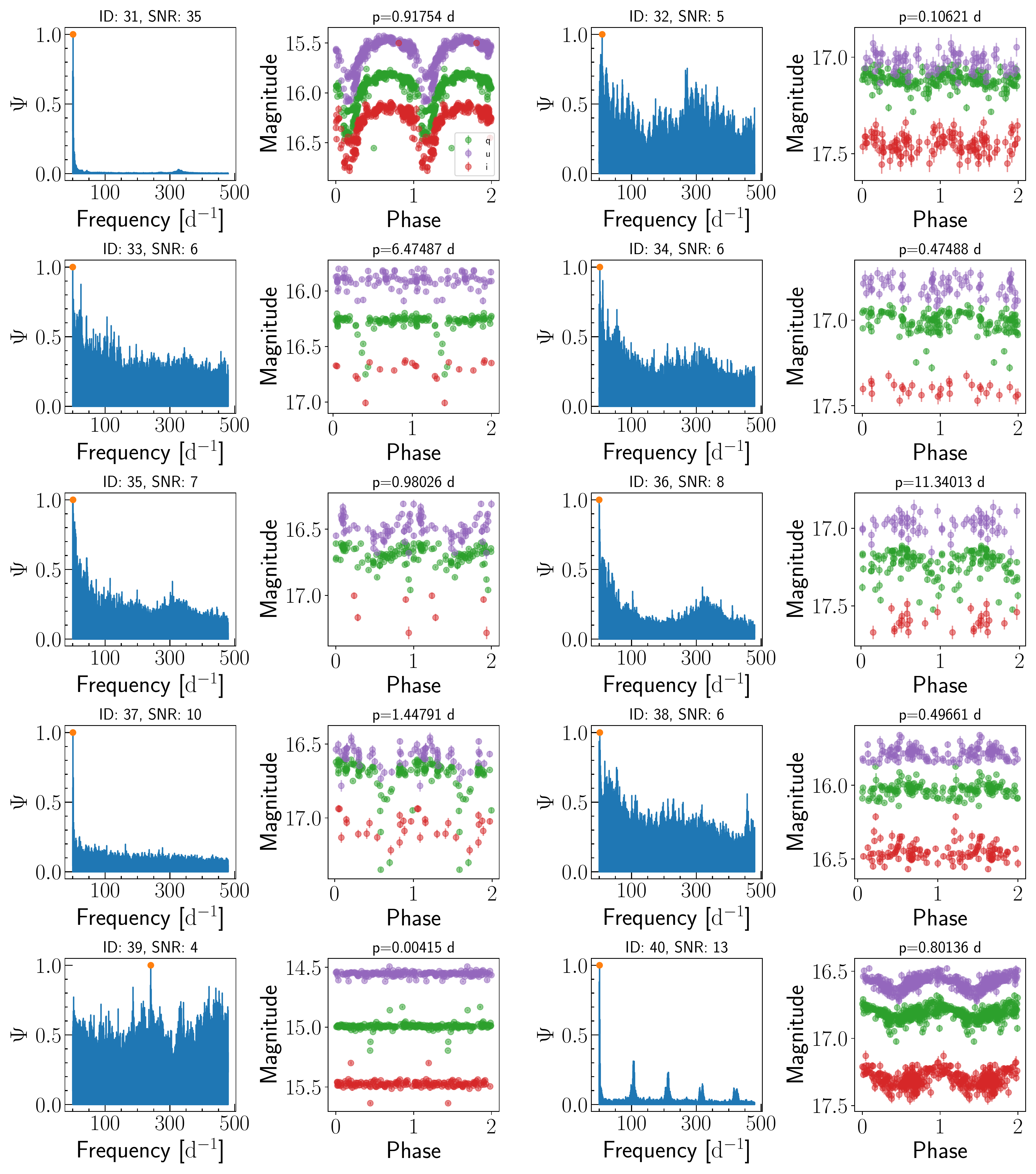}
  \caption{Periodograms and phase diagrams of MeerLICHT's candidates flagged as VARIABLE in \textit{Gaia} DR3.}
  \label{fig:gaia3}
  
\end{figure*}

\begin{figure*}
  \centering
  \includegraphics[trim={0cm 25cm 0cm 0cm},width=1\linewidth]{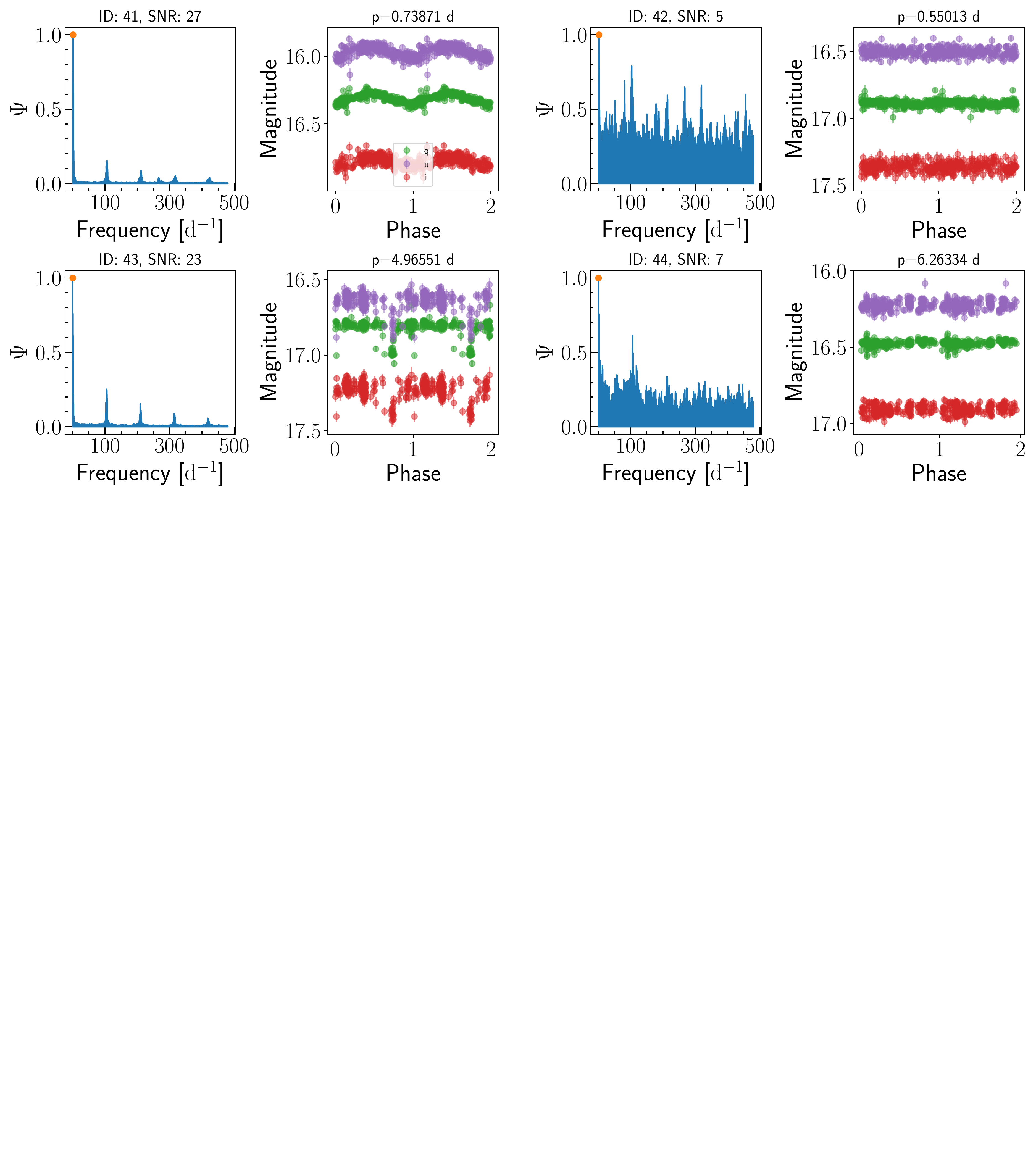}
  \caption{Periodograms and phase diagrams of MeerLICHT's candidates flagged as VARIABLE in \textit{Gaia} DR3.}
  \label{fig:gaia4}
\end{figure*}

This section describes sdB candidates identified already as Variables in \textit{Gaia} DR3 \citep{Gaia2022main}. 
These candidates are among the 610 sdB candidates discussed in Sects. \ref{sec:sdb_selection} and \ref{sec:lightcurves}. Running the frequency search algorithm (see Sect. \ref{sec:freq_application}) on these candidates allows us to derive the periods and possible variability classes based on their phase light curve shapes. The same approaches described in Sect. \ref{sec:freq_application} are applied to derive their periods using a one-term model. Visual inspection of the candidate phase light curves (Figs.~\ref{fig:gaia0}$-$\ref{fig:gaia4}) reveals that 13 of them are EBs and 4 reveal sinusoidal variability.  A summary of their photometric measurements and periods can be found in Table ~\ref{tab:gaia_candidate_list} by referring to the ID value on top of each plot. This table contains the coordinates, averaged magnitude in the three band filters ($u,q,i$),  number of observations in each filter (N$q$, N$u$, N$i$), period, S/N, and a preliminary classification of the variability type of each candidate. We note that most of the candidates have no assigned class as their variability type is not obvious from the light curve shapes.
\onecolumn


\clearpage{}

\begin{table}[h!]
    \centering
    \caption{MeerLICHT candidates flagged as VARIABLE in \textit{Gaia} DR3. }\label{tab:gaia_candidate_list} 
\begin{tabular}{cccccccccccc}
\toprule
ID & RA & Dec. & $q$ & $u$ & $i$ & N$q$ & N$u$ & N$i$ & Period & S/N & Class\\   \smallskip
& (deg.) & (deg.) & (mag.) & (mag.) & (mag.) & & & & (day) & \\ \midrule
1 & 72.98766 & -68.51423 & 16.71 & 16.28 & 17.11 & 314 & 198 & 184 & 273.9462 & 30 & $-$ \\
2 & 80.82555 & -68.25012 & 15.42 & 14.97 & 15.89 & 85 & 80 & 41 & 1.57 & 21 & EB \\
3 & 81.80675 & -67.21868 & 17.21 & 16.87 & 17.66 & 79 & 51 & 45 & 0.6982 & 5 & EB \\
4 & 83.92374 & -66.89975 & 16.87 & 16.36 & 17.36 & 79 & 67 & 60 & 0.003 & 5 & $-$ \\
5 & 74.36943 & -66.86057 & 15.48 & 15.11 & 15.93 & 137 & 114 & 83 & 0.8584 & 11 & Sinusoidal \\
6 & 74.09615 & -66.83624 & 16.09 & 15.78 & 16.54 & 135 & 113 & 98 & 0.992 & 13 & $-$ \\
7 & 73.66313 & -66.83476 & 17.27 & 17.04 & 17.65 & 128 & 40 & 38 & 0.125 & 7 & $-$ \\
8 & 74.21588 & -66.66547 & 16.42 & 16.02 & 16.91 & 133 & 108 & 93 & 0.2489 & 8 & $-$ \\
9 & 73.19782 & -66.6623 & 16.75 & 16.41 & 17.22 & 110 & 73 & 32 & 1.0076 & 6 & $-$ \\
10 & 17.21972 & -72.31037 & 16.41 & 16.07 & 16.87 & 194 & 213 & 44 & 0.7256 & 21 & EB \\
11 & 81.74164 & -66.34624 & 16.5 & 16.12 & 16.93 & 95 & 85 & 78 & 0.8988 & 5 & $-$ \\
12 & 80.52864 & -66.15882 & 14.93 & 14.64 & 15.32 & 107 & 94 & 88 & 3.3202 & 32 & EB \\
13 & 82.14073 & -66.14558 & 16.15 & 15.83 & 16.6 & 107 & 89 & 82 & 0.715 & 30 & EB \\
14 & 79.26889 & -66.06711 & 15.15 & 14.72 & 15.63 & 82 & 96 & 59 & 0.5021 & 9 & $-$ \\
15 & 83.53014 & -67.49373 & 15.91 & 15.47 & 16.37 & 82 & 83 & 76 & 0.7408 & 12 & EB \\
16 & 203.73375 & -33.38689 & 13.99 & 13.5 & 14.52 & 100 & 88 & 74 & 0.4986 & 5 & $-$ \\
17 & 76.77216 & -67.96542 & 15.18 & 14.89 & 15.63 & 260 & 266 & 246 & 1.7562 & 15 & $-$ \\
18 & 76.91421 & -67.67346 & 16.64 & 16.35 & 17.1 & 281 & 217 & 205 & 0.4547 & 26 & Sinusoidal \\
19 & 75.25306 & -67.18034 & 16.58 & 16.35 & 16.99 & 108 & 102 & 91 & 0.7073 & 14 & EB \\
20 & 76.24324 & -66.73309 & 17.28 & 16.96 & 17.71 & 132 & 37 & 41 & 12.3481 & 6 & $-$ \\
21 & 74.54947 & -66.61895 & 16.11 & 15.66 & 16.61 & 135 & 109 & 99 & 0.2102 & 4 & $-$ \\
22 & 28.52677 & -74.07146 & 17.1 & $-$ & 17.58 & 44 & 0 & 1 & 0.101 & 4 & $-$ \\
23 & 28.10981 & -73.99584 & 17.46 & $-$ & $-$ & 44 & 0 & 0 & 0.0024 & 3 & $-$ \\
24 & 82.14495 & -70.79937 & 16.12 & 15.76 & 16.44 & 78 & 64 & 62 & 3.3483 & 5 & $-$ \\
25 & 79.75311 & -67.99859 & 16 & 15.59 & 16.45 & 284 & 259 & 211 & 245.8972 & 10 & $-$ \\
26 & 81.4855 & -67.97021 & 16.32 & 15.87 & 16.76 & 75 & 76 & 69 & 0.318 & 15 & EB \\
27 & 79.60767 & -67.92812 & 15.92 & 15.7 & 16.3 & 287 & 250 & 245 & 256.3609 & 9 & $-$ \\
28 & 79.64751 & -67.85666 & 15.94 & 15.57 & 16.41 & 288 & 264 & 250 & 0.709 & 63 & EB \\
29 & 203.73374 & -33.38689 & 13.99 & 13.5 & 14.52 & 100 & 88 & 74 & 0.4986 & 5 & $-$ \\
30 & 74.65939 & -69.49231 & 16.43 & 16.16 & 16.78 & 314 & 243 & 226 & 0.9993 & 22 & EB?\\
31 & 73.64486 & -69.19692 & 15.9 & 15.55 & 16.23 & 319 & 280 & 272 & 0.9175 & 35 & EB \\
32 & 80.52673 & -66.43712 & 17.12 & 17.03 & 17.45 & 108 & 44 & 63 & 0.1062 & 4 & $-$ \\
33 & 73.51213 & -70.98084 & 16.26 & 15.9 & 16.68 & 79 & 58 & 11 & 6.4749 & 5 & EB \\
34 & 74.17132 & -70.86893 & 17 & 16.81 & 17.4 & 85 & 37 & 15 & 0.4749 & 6 & $-$ \\
35 & 73.47833 & -70.86706 & 16.69 & 16.5 & 17.1 & 76 & 52 & 4 & 0.9803 & 6 & $-$ \\
36 & 74.37838 & -70.85005 & 17.2 & 16.97 & 17.61 & 80 & 34 & 15 & 11.3401 & 7 & $-$ \\
37 & 73.85634 & -70.78173 & 16.67 & 16.57 & 17.04 & 78 & 38 & 16 & 1.4479 & 10 & EB \\
38 & 76.1173 & -70.20635 & 16.03 & 15.78 & 16.46 & 91 & 76 & 65 & 0.4966 & 5 & $-$ \\
39 & 73.06925 & -66.55055 & 14.99 & 14.55 & 15.47 & 123 & 103 & 85 & 0.0041 & 4 & $-$ \\
40 & 15.19798 & -72.09575 & 16.81 & 16.58 & 17.31 & 285 & 218 & 194 & 0.8014 & 13 & Sinusoidal \\
41 & 12.88436 & -71.66993 & 16.3 & 15.96 & 16.77 & 286 & 256 & 220 & 0.7387 & 26 & Sinusoidal \\
42 & 13.26228 & -71.62152 & 16.89 & 16.51 & 17.36 & 273 & 194 & 174 & 0.5501 & 5 & $-$ \\
43 & 15.64936 & -71.45853 & 16.8 & 16.64 & 17.22 & 243 & 213 & 206 & 4.9655 & 23 & EB \\
44 & 14.77471 & -71.19599 & 16.47 & 16.23 & 16.91 & 285 & 254 & 244 & 6.2633 & 6 & $-$ \\ 
\bottomrule \\
\end{tabular}
\\ \vspace{10pt}
\textbf{ Notes.}\justifying The corresponding periodograms and phase diagrams of each candidate are shown in Figs. \ref{fig:gaia0}$-$\ref{fig:gaia4}. The listed IDs are the same as those indicated in these figures. The $q, u,$ and $i$ columns represent the median of the magnitudes in the $q-, u-, $ and $i-$band filters, respectively, while N$q$, N$u$, and N$i$ correspond to the number of observations in the $q-, u-, $ and $i-$band filters, respectively. The last column, Class, denotes the preliminary variability classification of the candidates; those with unclear variability in their phased lightcurves were left unclassified.
\end{table}

\clearpage{}
\section{Candidates' variability indices plots and full catalogue}\label{appendix:B}

\begin{figure*}[h!]
  \centering
  \includegraphics[trim={0cm 0cm 0cm 0cm},width=1\linewidth]{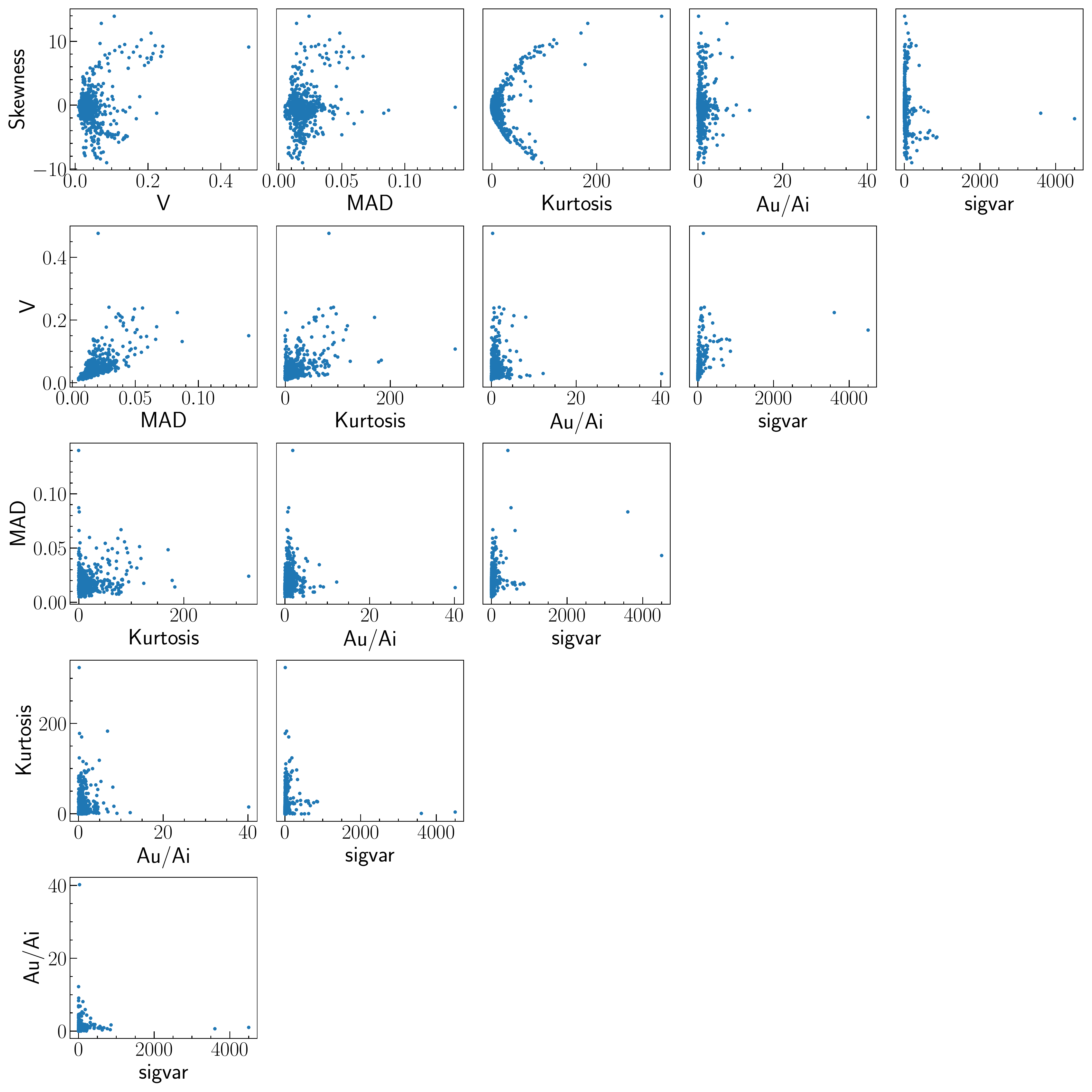}
  \caption{Plots of the 610 sdB candidates' variability indices. Au/Ai is the phase-folded models' amplitude ratio between the $u$- and $i$-band filters for the dominant frequency, whereas sigvar is the significance of variability (also known as the reduced weighted $\chi^2$) computed from candidates' fluxes in the $q$-band filter. The skewness and kurtosis \citep{Friedrich1997} are obtained from the $q$-band fluxes, while the MAD and magnitude of variability (V), defined in Sect. \ref{sec:freq_application}, are derived from the $q$-band magnitudes.}
  \label{fig:var_indices_plots1}
\end{figure*}

\begin{figure*}
  \centering
  \includegraphics[trim={0cm 0cm 0cm 0cm},width=1\linewidth]{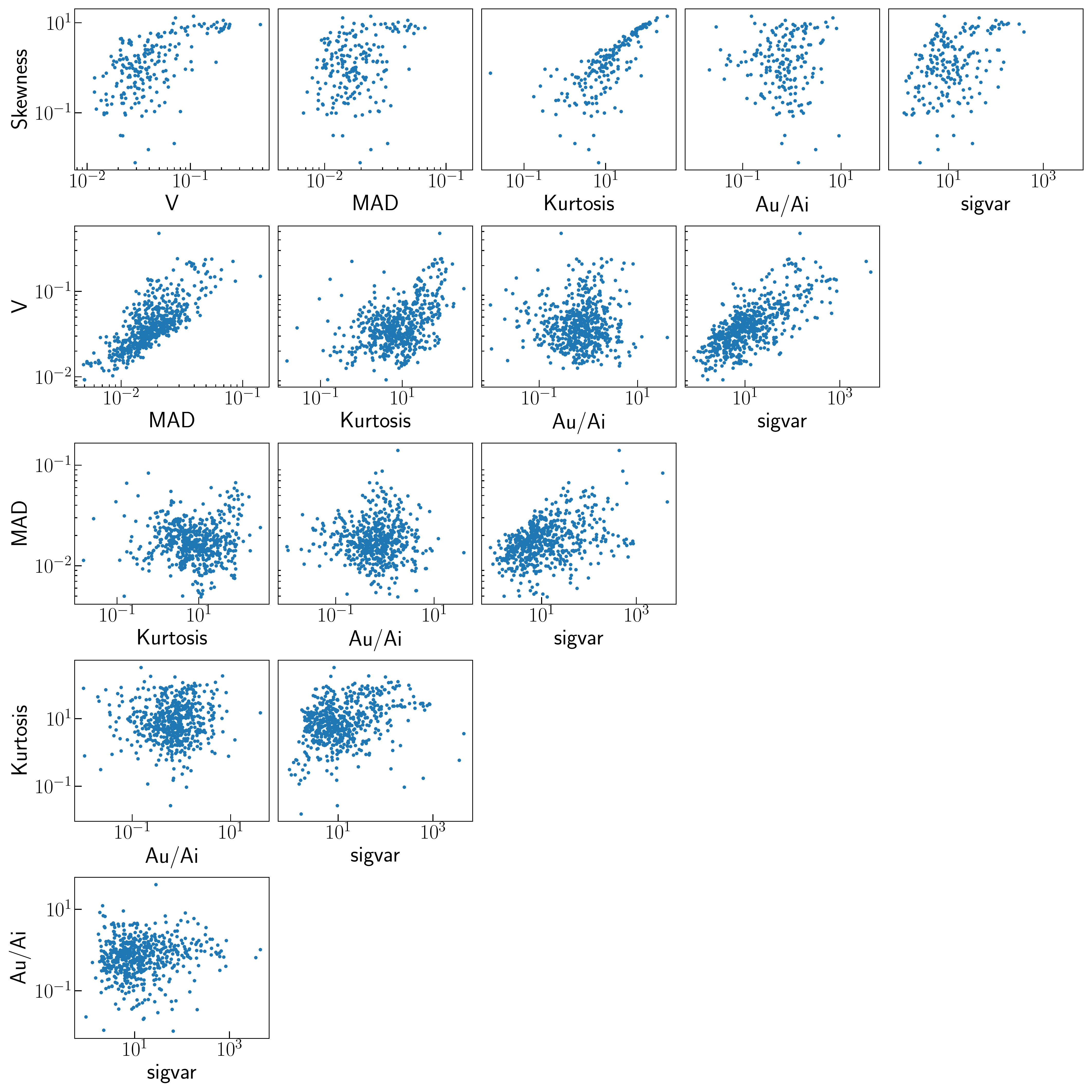}
  \caption{Same as Fig. \ref{fig:var_indices_plots1}, but the x- and y-axis are represented on a logarithmic scale.}
  \label{fig:var_indices_plots2}
\end{figure*}

\end{appendix}

\end{document}